\documentclass[fleqn,usenatbib]{mnras}

\usepackage{newtxtext,newtxmath}

\usepackage[T1]{fontenc}

\DeclareRobustCommand{\VAN}[3]{#2}
\let\VANthebibliography\thebibliography
\def\thebibliography{\DeclareRobustCommand{\VAN}[3]{##3}\VANthebibliography}

\usepackage{graphicx}	
\usepackage{amsmath}	
\usepackage{textcomp}	
\usepackage[dvipsnames]{xcolor}
\usepackage{tikz}
\usetikzlibrary{arrows.meta,
                 shapes,
                chains,
                positioning,
                shapes.geometric}
\tikzstyle{arrow}=[draw, -latex]

\newcommand{\appropto}{\mathrel{\vcenter{
  \offinterlineskip\halign{\hfil$##$\cr
    \propto\cr\noalign{\kern2pt}\sim\cr\noalign{\kern-2pt}}}}}

\newcommand{\cs}{c_\mathrm{s}}

\newcommand{\Mt}{M_\mathrm{b}}

\newcommand{\ab}{a_\mathrm{b}}
\newcommand{\qb}{q_\mathrm{b}}
\newcommand{\eb}{e_\mathrm{b}}
\newcommand{\Pb}{P_\mathrm{b}}

\newcommand{\OmegaP}{\Omega_\mathrm{P}}
\newcommand{\Omegab}{\Omega_\mathrm{b}}
\newcommand{\OmegaK}{\Omega_\mathrm{K}}

\newcommand{\de}{\mathrm{d}}

\newcommand{\td}[2]{\frac{\de #1}{\de #2}}

\newcommand{\azav}[1]{\left\langle #1 \right\rangle_\phi}

\newcommand{\const}{\mathrm{const.}}

\newcommand{\I}{\mathrm{i}}

\newcommand{\Rcav}{R_\mathrm{cav}} 

\newcommand{\Tex}{\frac{\de {T_\mathrm{ex}}}{\de R}}
\newcommand{\Texm}{\frac{\de {T_\mathrm{ex,m}}}{\de R}}
\newcommand{\Texs}{\de T_\mathrm{ex}/\de R}
\newcommand{\Texms}{\de T_\mathrm{ex,m}/\de R}

\newcommand{\Tdeps}{\de T_\mathrm{dep}/\de R}

\defcitealias{Goodman2001}{GR01}
\defcitealias{Rafikov2002}{Rafikov 2002}
\defcitealias{Cimerman2021}{Paper I}

\title[Torque in circumbinary discs]
{Gravitational torque in circumbinary discs: global radial oscillations}

\author[N. P. Cimerman, R. R. Rafikov]{
Nicolas P. Cimerman$^{1}$,
Roman R. Rafikov$^{1,2}$\thanks{E-mail: rrr@damtp.cam.ac.uk (RRR)}
\\
$^{1}$Department of Applied Mathematics and Theoretical Physics, University of Cambridge, Wilberforce Road, Cambridge CB3 0WA, UK\\
$^{2}$Institute for Advanced Study, Einstein Drive, Princeton, NJ 08540, USA
}

\date{Accepted XXX. Received YYY; in original form ZZZ}

\pubyear{2022}

\begin{document}

\label{firstpage}
\pagerange{\pageref{firstpage}--\pageref{lastpage}}
\maketitle

\begin{abstract}
Circumbinary discs (CBDs) arise in many astrophysical settings, including young stellar binaries and supermassive black hole binaries. Their structure is mediated by gravitational torques exerted on the disc by the central binary. The spatial distribution of the binary torque density (so-called excitation torque density) in CBDs is known to feature global large-amplitude, quasi-periodic oscillations, which are often interpreted in terms of the local resonant Lindblad torques. Here we investigate the nature of these torque oscillations using 2D, inviscid hydrodynamic simulations and theoretical calculations. We show that torque oscillations arise due to the gravitational coupling of the binary potential to the density waves launched near the inner cavity and freely propagating out in the disc. We provide analytical predictions for the radial periodicity of the torque density oscillations and verify them with simulations, showing that disc sound speed and the multiplicity of the density wave spiral arms are the key factors setting the radial structure of the oscillations. Resonant Lindblad torques play no direct role in determining the radial structure and periodicity of the torque oscillations and manifest themselves only by driving the density waves in the disc. We also find that vortices forming at the inner edge of the disc can provide a non-trivial contribution to the angular momentum transport in the CBD. Our results can be applied to understanding torque behaviour in other settings, e.g. discs in cataclysmic variables and X-ray binaries.
\end{abstract}





\section{Introduction}
\label{sec:intro}


Gaseous circumbinary discs (CBDs) naturally arise in a number of astrophysical settings: formation of young binary stars \citep{Dutrey2016,Yang2017}, circumbinary planet formation \citep{Mesch2012,Pelup2013,Silsbee2015}, post-AGB  \citep{deRuy2006} and X-ray \citep{Blun2008} binaries and supermassive black hole (SMBH) binaries \citep{Armitage2002,MacFadyen2008}. The ubiquity of CBDs makes understanding of their properties and dynamics very important, and a number of recent theoretical, numerical and observational studies have explored them. 

A characteristic feature of CBDs is the cavity that forms at the disc centre where the binary orbits \citep{Art1996,Armitage2002}. The radius of the cavity $\Rcav$ is typically found to lie in the interval $(2-5)\ab$, where $\ab$ is the binary semi-major axis, depending on the binary mass ratio, eccentricity, as well as the disc aspect ratio and viscosity \citep{Pelup2013,Ragusa2020,Hirsh2020,Ditt2022}.  The cavity arises due to the gravitational (tidal) torques exerted by the binary on the disc \citep{Art1994}, similar to the origin of gaps carved out by massive planets in protoplanetary discs \citep{Pap1984}. The value of $\Rcav$ is set by the competition between the injection of angular momentum into the disc by the binary torque and the internal (viscous) disc stresses. 

It is important to recognize that tidal binary-disc coupling operates as a {\it nonlocal, two-stage} process \citep{Lunine1982,Greenberg1983,Goldreich1989,RP12,PR12}: first, the torque due to the non-axisymmetric component of the binary potential excites a density wave in the disc, adding angular momentum to the wave at the rate (per unit radius $R$) given by the {\it excitation} torque density $\Texs$, defined explicitly in Section \ref{sec:methods}. After propagating over some distance, the wave dissipates (via linear or nonlinear processes) ultimately transferring angular momentum to the disc fluid at the rate (per unit $R$) given by the {\it deposition} torque density $\Tdeps$, which is different from $\Texs$. It is set the balance of the deposition torque density and the internal stresses that sets the value of $\Rcav$. However, it is clear that the excitation torque density is still a crucial part of the disc dynamics: after all, $\Texs$ is one of the factors determining the spatial distribution of $\Tdeps$ (see Section \ref{sec:res-torque}). For this reason, understanding the behaviour (radial profile) of $\Texs$ is very important for clarifying the properties of CBDs. 

\begin{figure}
	\begin{center}
	\includegraphics[width=0.49\textwidth]{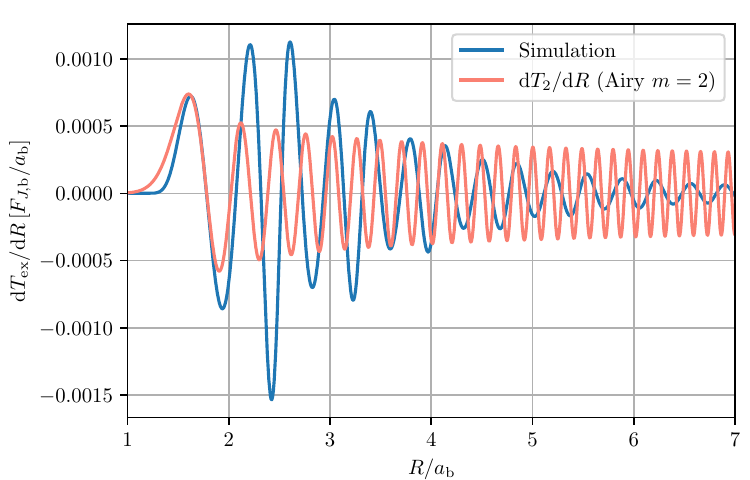}
	\caption{
    Radial profile of the torque density $\Texs$ extracted from a simulation (blue) of a mass ratio $\qb=10^{-3}$ binary with a circumbinary disc, see Section \ref{sec:setup} for the setup and details. A globally isothermal equation of state with the aspect ratio $h_0=0.1$ (at $R=\ab$, see Eq. (\ref{eq:h})) is adopted here. Note the oscillatory, quasi-periodic behaviour of $\Texs$.  The salmon curve shows the resonant torque $\de T_2/\de R$ (depending on $R$ via the Airy function) given by Eq. (\ref{eq:res-T}), with the amplitude chosen arbitrarily to match the first peak of $\Texs$; see Section \ref{sec:res-torque} for details. 
 }
	\label{fig:Airy}
	\end{center}
\end{figure}

A number of past studies explored $\Texs$ as a function of $R$ in simulations of CBDs for a variety of binary and disc parameters and physical assumptions \citep{MacFadyen2008,Farris2011,Farris2014,Shi2012,Noble2012,Noble2021,Roedig2012,DOrazio2013,Zil2015,Miranda2017,Dempsey2020,Tiede2022,WangBai2022a,Wang2022,Mahesh2023}, generally agreeing on a set of common features. First, $\Texs$ is quite small inside the cavity, which is easily explained by the low surface density of gas in that region. Second, starting from $R\sim\Rcav$ and extending far in the disc, $\Texs$ features prominent quasi-periodic oscillations in $R$, attaining both positive and negative values. The amplitude of these oscillations decays with the distance but they can often be traced out to $R\gtrsim (5-10)\ab$.  Third, the radial integral of $\Texs$ converges to a positive value as $R\to \infty$, indicating that the gravitational torque due to the binary drives an {\it outward} flow of angular momentum through the CBD. These features are illustrated in Fig.  \ref{fig:Airy}, where we show $\Texs$ (blue curve) extracted from one of our simulations (described in detail in Section \ref{sec:results}) of a CBD around a binary with a mass ratio $\qb=10^{-3}$.   

Understanding the spatial structure of $\Texs$ and, more specifically, the nature of its global radial oscillations, is the primary goal of our present work. Starting with the seminal work of \citet{MacFadyen2008}, the standard explanation for the excitation torque density oscillations has been to associate $\Texs$ with the azimuthal $m=2$ component of the local torque density $\de T_2/\de R$ excited in the vicinity of $3:2$ mean motion resonance with the binary. The use of the $m=2$ component of the torque density is naturally motivated by the ubiquitous presence of two prominent spiral arms in many CBD simulations \citep{MacFadyen2008}. The radial dependence enters the resonant torque eigenfunction via the argument of an Airy function \citep{Meyer1987}, which has oscillatory behaviour; see Fig. \ref{fig:Airy} and Eq. (\ref{eq:res-T}) for the explicit representation of $\de T_2/\de R$. Because of that, it was suggestive to relate $\de T_2/\de R$ to the radial oscillations of $\Texs$. This explanation of $\Texs$ oscillations has also been invoked in \citet{Farris2011} and \citet{Shi2012}.

However, it has been noted already in \citet{MacFadyen2008} that the radial structure of $\de T_2/\de R$ provides a rather poor fit to the spatial distribution of $\Texs$ on global scales. Indeed, because of the Airy function dependence, the radial scale of $\de T_2/\de R$ oscillation decreases with $R$ for $R\gtrsim \ab$, see salmon curve in Fig.  \ref{fig:Airy} and Section \ref{sec:resonant-torque}. This behaviour is not supported by simulations, which generally show a different evolution of the radial period of $\Texs$, as evident from Fig.  \ref{fig:Airy}. 

In this study we provide an alternative explanation for $\Texs$ oscillations, which successfully reproduces their spatial variation. Our approach is closely related to the recent study \citep{Cimerman2023b} of the so-called torque wiggles in disc-planet interaction --- small amplitude, quasi-periodic oscillations of the excitation torque density arising far from the planet. That work, focusing on the regimes of linear coupling (i.e. $q_\mathrm{b}\to 0$) and a smooth underlying disc around the planet (i.e. no gap), has shown that the global oscillations of $\Texs$ result from the gravitational coupling of the binary to the freely-propagating planet-driven density wave. Although in the binary case the underlying disc structure is very different due to the presence of a central cavity, and $\qb$ can be as high as unity, we show that the nature of $\Texs$ oscillations in CBDs is pretty much the same as in \citet{Cimerman2023b}. 

In the course of this study we also found that for $\qb$ approaching unity the spatial structure of the surface density perturbation and the angular momentum budget in the inner disc exhibit curious anomalies, which are caused by the vortex-driven modes \citep{Coleman2022a} excited by the vortices forming at the cavity edge. These vortices may be intimately related to lumps --- non-axisymmetric overdensities that are frequently found to form at the cavity edge in long-term simulations of CBDs \citep{Shi2012,Noble2012}.

Our work is organized as follows. After describing our physical and numerical setup and methodology in Section \ref{sec:setup}, we present our results on the torque density behaviour in Section \ref{sec:results}, including theoretical understanding of the origin of $\Texs$ oscillations in Sections \ref{sec:torque-origin}\&\ref{sec:torque-analysis}. In Section \ref{sec:vortices} we describe the emergence of vortices in our simulations, their contribution to the angular momentum transport in the disc and their possible role in the lump formation at the cavity edge. We provide a discussion of our results in Section \ref{sec:discuss}, including the role of resonant torques (Section \ref{sec:resonant-torque}) and comparison with the planetary case (Section \ref{sec:planet}), and summarize our findings in Section \ref{sec:conc}.


\section{Problem setup and methodology}
\label{sec:setup}

 
To elucidate the nature of the torque density oscillations we design a particular CBD setup that makes it easy to reveal their nature using numerical simulations. 

First, since it takes only a short amount of time for the $\Texs$ oscillations to develop in the CBD --- basically, the sound crossing time across the simulation domain --- we do not run most of our simulations for too long, see Section \ref{sec:disc_model}. This allows us to avoid the complications that arise in long-term CBD simulations: development of the eccentricity in the inner disc \citep{MacFadyen2008,Farris2014}, emergence of the lump at the cavity edge \citep{Shi2012,Noble2012}, etc. 

Second, since we are interested only in the gravitational torques in the CBD outside the cavity edge, we do not need to resolve the physics of what is happening inside the cavity particularly accurately. For that reason we use a numerical setup with an excised central region (Section \ref{sec:disc_model}). 

Third, to reduce the number of physical processes that may complicate the interpretation of our results, in this study we mainly focus on inviscid CBDs (although we also describe some simulations with non-zero viscosity in Section \ref{sec:lumps}), which did not receive much attention so far. Also, to lower the number of relevant parameters we use simple thermodynamic assumptions, as outlined in Section \ref{sec:physics}.


\subsection{Physical setup}
\label{sec:physics}


We consider a binary of two point-masses $M_1 \geq M_2$, with total mass $\Mt = M_1 + M_2$ and introduce the mass-ratio $\qb = M_2/M_1 \leq 1$. The binary orbit is circular and fixed in time, i.e. $\ab=$ const and binary eccentricity $\eb = $ const $ = 0$. We work in the inertial (i.e. non-rotating) frame, with the origin of the polar coordinates $(R,\phi)$ coinciding with the centre of mass of the binary (note the difference with \citet{Cimerman2023b} who worked in a frame centred on and co-moving with one of the components of the binary). The binary orbital frequency is $\Omegab = \left(G\Mt/\ab^3\right)^{1/2}$, such that the coordinates of the (circular) binary components are given by ${\bf R}_1=(R_1, \phi_1) = (\qb \ab/(\qb + 1), \Omegab t)$ and ${\bf R}_2=(R_2, \phi_2) = (\ab/(\qb + 1), \Omegab t + \pi)$.

The gravitational potential of the binary is:
\begin{align}
	\nonumber
	\Phi_\mathrm{b}
	= \Phi_1 + \Phi_2,
	\label{eq:phib}
\end{align}
where the individual contributions are
\begin{align}
	\Phi_i(R,\phi) = -\frac{G M_i}{\left|{\bf R}-{\bf R}_i\right|}.
\end{align}
We do not soften the potential, since the binary components never enter the simulation domain (see below).

The binary orbit lies within the cavity of a two-dimensional, coplanar circumbinary disc. We neglect self-gravity of the disc gas and do not account for the back-reaction it may have on the binary. 

Our treatment of disc thermodynamics generally assumes a locally isothermal equation of state (EoS) $P = \cs^2(R) \Sigma$, where $P$,  $\Sigma$ and $\cs$ are the (vertically integrated) pressure, surface density and sound speed of disc gas. The sound speed is set using the power-law temperature profile $T(R)\propto R^{-q}$. For $R \gg \ab$, where the local $\OmegaK(R) \approx \sqrt{G\Mt/R^3}$, the disc scale-height is $H = \cs / \OmegaK$ and the disc aspect ratio $h=H/R$ can be expressed as
\begin{align}
h(R)=h_0\left(\ab/R\right)^{(q-1)/2},
\label{eq:h}
\end{align}
where $h_0$ is the nominal value of $h$ at $R=\ab$.

However, in most of our calculations (except in Section \ref{sec:discuss}), we use the {\it globally} isothermal EoS with radially independent $T$ and $\cs$ (i.e. $q=0$). This is done partly for simplicity, since in this case the constant sound speed $\cs$ (or, equivalently, the value of $h_0$) becomes the only physical parameter characterizing disc thermodynamics. Another reason for choosing this EoS is that it preserves the angular momentum flux carried by the density waves in the disc. As was shown recently by \citet{Miranda2019II,Miranda2020I}, wave angular momentum conservation breaks down for other, more general thermodynamic setups, e.g. a locally isothermal EoS with radially varying $\cs$. Avoiding this outcome somewhat simplifies the interpretation of our results.

Our choice of the EoS with $q=0$ has a slight disadvantage that the disc aspect-ratio $h = H/R \propto R^{1/2}$, see equation (\ref{eq:h}), i.e. the disc is flared. However, in our analysis we use the data only for $R \leq 8 \ab$, for which the maximum aspect ratio is $h(R=8\ab) \simeq 0.28$; the disc flaring beyond that point is not worrisome. 

As mentioned earlier, most of our runs do not include explicit viscosity. However, we do consider viscosity characterized via a dimensionless $\alpha$-parameter \citep{Shakura1973} in some illustrative simulations in Section \ref{sec:lumps}.


\subsection{Numerical setup}
\label{sec:disc_model}


Our numerical setup is similar to the one used in \citet{Cimerman2023b}, with the important difference that now we work in the inertial polar frame centred on the system barycentre. We perform our hydrodynamic simulations using Athena++ 
\citep{Athenapp2020}. The code uses a Godunov scheme to solve the hydrodynamic equations in a conservative form fully accounting for the binary potential $\Phi_\mathrm{b}$.
We use Roe's approximate Riemann solver, second-order Runge-Kutta time-stepping and second-order spatial reconstruction. 

We run most of our simulations for 100 binary orbits, which is a relatively short time. We find this duration to be sufficient for the $\Texs$ pattern to fully develop (and for vortices to appear in high-$\qb$ runs), so that we can perform a meaningful averaging of the fluid variables. Apart from lowering the numerical cost of the runs, such a short duration does not allow gas streamlines near the cavity edge to develop substantial eccentricity (which often happens in long-term CBD simulations) simplifying our analysis. In Section \ref{sec:comp_prev} we discuss how our results could be modified in longer term simulations.

\begin{figure}
	\begin{center}
	\includegraphics[width=0.49\textwidth]{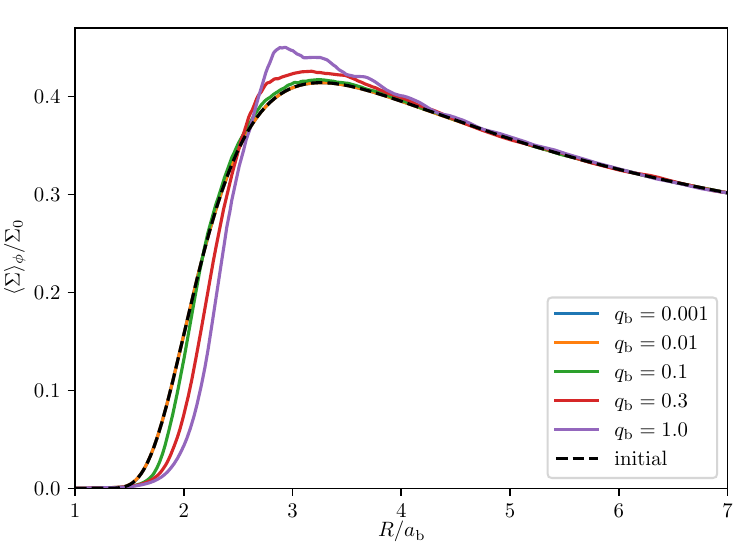}
	\caption{
		Radial profile of the azimuthally-averaged surface density $\azav{\Sigma}$ for different values of $\qb$ used in this work. We show $\azav{\Sigma}$ time-averaged from $t = 40 \Pb$ to $50 \Pb$ and sampled ten times per binary orbit $\Pb$, as well as the initial surface density profile (black dashed curve) given by the equation (\ref{eq:Sigma0}).
	}
	\label{fig:sig_r}
	\end{center}
\end{figure}


\subsubsection{Initial conditions}
\label{sec:ICs}

Our initial setup for the CBD includes a well-developed cavity, i.e. we do not track its formation. We follow the prescription of \citet{Munoz2019} for the radial profile of a cavity in the CBD and adopt the following axisymmetric surface density distribution at $t=0$, illustrated in Fig.  \ref{fig:sig_r}:
\begin{align}
	\Sigma(R,t=0) & = \Sigma_0 \left( \frac{R}{\ab} \right)^{-1/2}
	\left[ 1 - \frac{3}{4} \sqrt{\frac{\ab}{R}} \right] 
 \nonumber\\
 &\times
	\exp \left[-\left( \frac{R}{R_\mathrm{cav}} \right)^{-4} + \left(
		\frac{R}{R_\mathrm{cav}} \right)^{-2} \right],
  \label{eq:Sigma0}
\end{align}
where $R_\mathrm{cav} = 2.45 \ab$; the maximum of $\Sigma(R,t=0)$ is about $0.41\Sigma_0$ and is at $R\approx 3.26\ab$.  A density floor of $\Sigma_\mathrm{fl} = 10^{-9} \Sigma_0$ is used to avoid numerical instability in the cavity region.  The prescription (\ref{eq:Sigma0}) for the cavity shape is based on the steady-state profiles of $\Sigma(R)$ attained in viscous simulations of $\qb = 1$ binaries \citep{Miranda2017}.  Since here we mainly focus of the inviscid case and often on lower $\qb<1$, we do not expect $\Sigma(R,t=0)$ to be in perfect equilibrium and $\Sigma$ might evolve somewhat during our runs. However, this is not going to affect our results for $\Texs$ and their interpretation, see Section \ref{sec:discuss}.

Since we consider inviscid discs, we initialize radial velocity of the flow as $u_{R}(R,t=0) = 0$. The azimuthal velocity $u_\phi(R,t=0) = \Omega(R) R$ in the globally isothrmal setup is initially given by the radial force balance with 
\begin{align}
	\Omega^2(R) = \OmegaK^2 \left[1 + 3Q \left(\frac{R}{\ab}\right)^{-2} \right] + \frac{\cs^2}{R} \td{\ln \Sigma}{R},
 \label{eq:Omega0}
\end{align}
where the factor $Q = \qb(1+\qb)^{-2}/4$ accounts for the quadrupolar correction to the axisymmetric component of the binary gravity.

The use of Eqs. (\ref{eq:Sigma0}) and (\ref{eq:Omega0}) minimizes the evolution of the background disc properties after the start of the simulation.


\subsubsection{Grid and boundary conditions}
\label{sec:BCs}

We use a numerical grid that extends from $R = \ab$ to $R = 70\ab$ in radius with logarithmic spacing and covers the full azimuthal angle $0 \leq \phi < 2 \pi$ uniformly. The number of grid cells is $N_R = 1400$ and $N_\phi = 2200$ cells in radial and azimuthal directions, respectively. The large outer radius allows us to minimize its impact on the dynamics in the inner disc; in our analysis we use only the data for $R\le 8\ab$.

The inner boundary condition (BC) is a 'diode' or 'outflow only' BC \citep[e.g.][]{Miranda2017}: the values of the fluid variables in the ghost cells are always set to achieve zero radial gradient for $\Sigma$ and $u_\phi$. The radial velocity in the ghost cells is set depending on its sign in the active cell: for $u_R < 0$, the same value is set in the ghost cell enabling free inflow; for $u_R > 0$ the ghost cell is set to $-u_R < 0$ to give zero net-flux of matter across the boundary and avoid mass injection into the simulation domain. 

The values of the fluid variables in the outer ghost cells are set constant and equal to the initial data. However our results are not sensitive to this choice, since the outer radius is set artificially large and density waves do not reach the boundary during the duration of the most runs we consider (so wave reflection is not an issue).

\begin{figure*}
	\begin{center}
	\includegraphics[width=\textwidth]{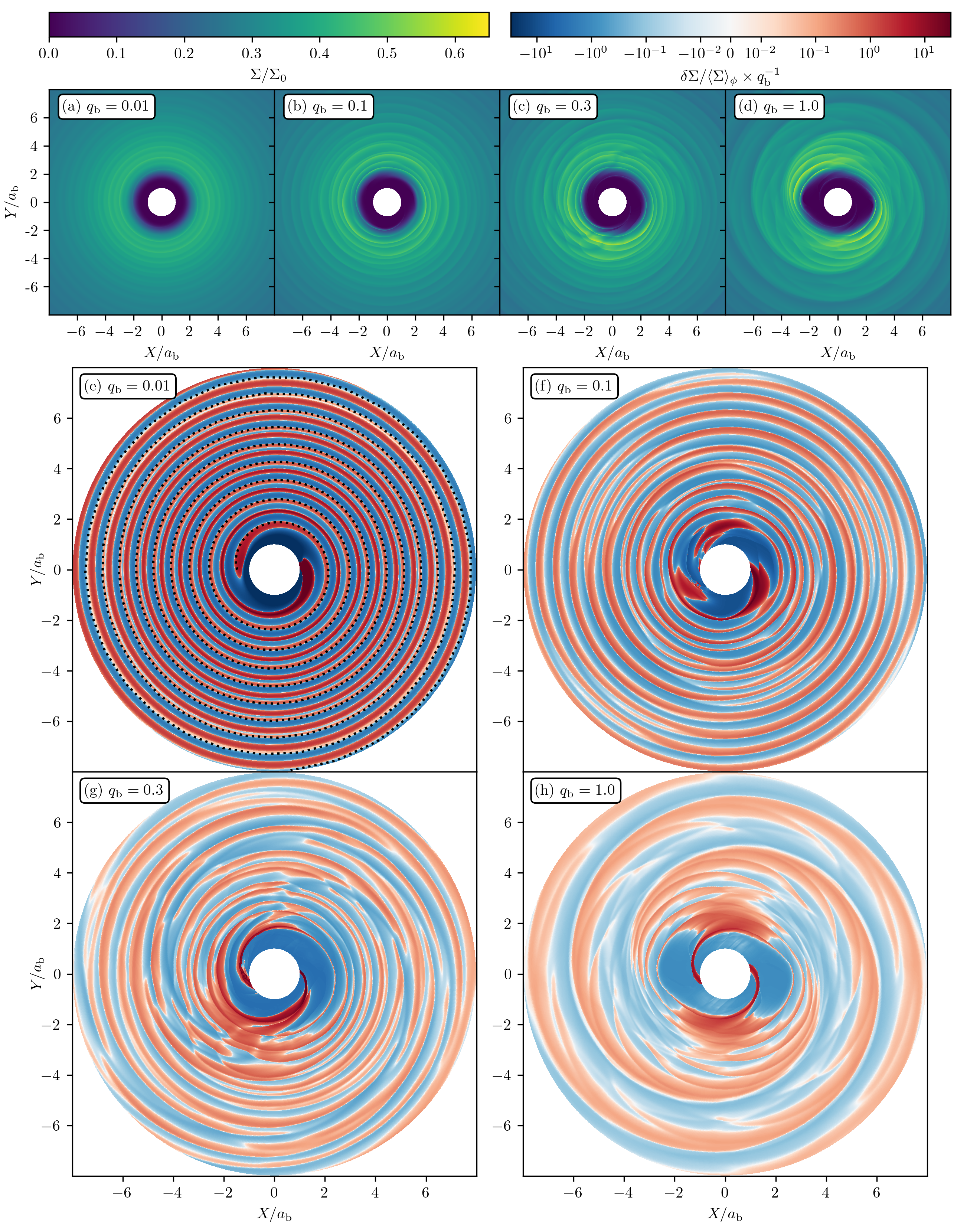}
        \vspace*{-0.1cm}
	\caption{
        2D maps at $t = 50 \Pb$ of the simulated (using globally isothermal EoS with $h_0=0.1$) surface density (top row) and \textit{relative} surface density perturbation $\Sigma/\azav{\Sigma} - 1$ (rest of figure), inversely scaled with the mass ratio $\qb$. The central white disc marks the excised region in which the binary orbits. At this time, the binary components are aligned with the $X$-direction. In panel (e)-(h), we do not show data for $R > 8\ab$, in order to make it easier to follow spiral features. The black dotted curve in panel (e) shows the shape of the density wave in the WKB approximation (Eq. \ref{eq:phi_m}) for $m=2$, $\OmegaP = \Omegab$, $R_\mathrm{OLR} = 1.45\ab$, $\phi_{2,0} = 0$ and $\Omega(R) = \azav{\Omega}(R)$ taken from the simulation. 
	}
	\label{fig:polar_sig}
	\end{center}
\end{figure*}


\subsection{Torque and angular momentum flux}
\label{sec:methods}


The excitation torque density due to the gravity of the binary acting on the disc is 
\begin{align}
\Tex &= -R\int_0^{2\pi}\Sigma(R,\phi)\left({\bf R}\times \nabla \Phi_\mathrm{b}\right)\,\de \phi	\nonumber\\
&=R\int_0^{2\pi}\Sigma(R,\phi)\sum\limits_{i=1,2}G M_i\frac{{\bf R}\times{\bf R}_i}{\left|{\bf R}-{\bf R}_i\right|^3}\,\de \phi.
\label{eq:Texs}
\end{align}
This formula is used to compute $\Texs$ in the barycentric frame \citep[cf.][]{Cimerman2023b} with $\Sigma(R,\phi)$ provided by our simulations.

We also define the angular momentum flux of the perturbation in the disc as
\begin{align}
	F_J(R) &= R^2 \int_0^{2\pi} \Sigma(R,\phi) u_R(R,\phi) \delta u_\phi(R,\phi) \,\de \phi,
	\label{eq:FJ}
\end{align}
where $\delta u_\phi = u_\phi - \tilde{u}_\phi$ and $\tilde{u}_\phi = \azav{u_\phi\Sigma}/\azav{\Sigma}$. In other words, we consider perturbations to the (generally time-dependent) density-weighted azimuthally-averaged $u_\phi$ as a background.

We extract azimuthal averages of surface density and momentum, as well as $\Texs$ and $F_J$ during run-time and save 100 snapshots of these processed data per orbit. This allows us to adequately average over potential time-variability of these quantities.


\subsection{Units}
\label{sec:units}


We use one binary orbital period $\Pb = 2\pi \Omegab^{-1}$ as a unit of time. Lengths are expressed in units of binary separation $\ab$ and surface density $\Sigma$ is normalized by $\Sigma_0$.

For the unit of $F_J$ we use a generalization of the (integrated) one-sided torque injected in the disc by a planet in linear disc-planet interaction \citep{GT80}:
\begin{align}
	F_{J,\mathrm{b}} = \qb^2 h_0^{-3} \Sigma_0 \ab^4 \Omegab^2.
	\label{eq:FJ_b}
\end{align}
This expression was also used in \citet{Cimerman2023b}, but now it features the reference surface density $\Sigma_0$ instead of $\Sigma$ at the location of the secondary (as in the planetary case) --- in the binary case the secondary is orbiting inside the cavity where $\Sigma \rightarrow 0$. Use of $F_{J,\mathrm{b}}$ provides us with a dimensionally suitable unit of $F_J$ that would reduce to the planetary case in the linear limit $\qb\to 0$ and without a cavity. Finally, the torque density $\Texs$ is measured in units of $F_{J,\mathrm{b}}/\ab$.

\begin{figure}
	\begin{center}
	\includegraphics[width=0.49\textwidth]{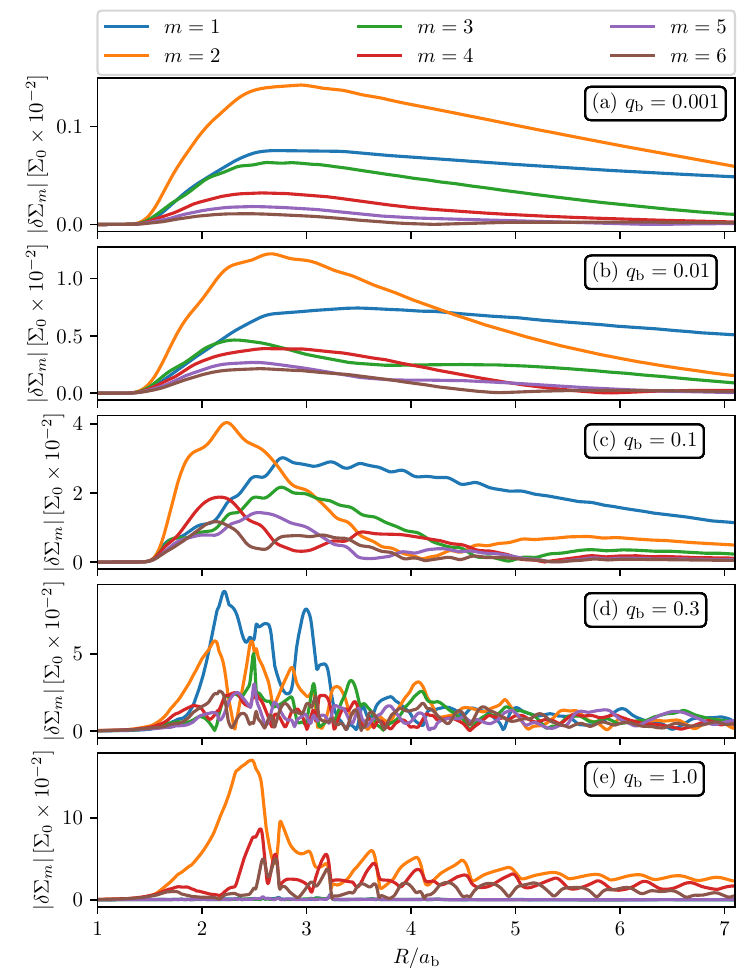}
	\caption{
		Radial profiles of the absolute magnitude of the Fourier coefficients $\delta \Sigma_m$ of the simulated density wave perturbation $\delta\Sigma$ at $t = 50\Pb$. Different panels show results for different values of $\qb$.
  }
	\label{fig:fourier_dsig}
	\end{center}
\end{figure}


\section{Results}
\label{sec:results}


Now we present the results of our globally isothermal CBD simulations with $h_0=0.1$ for several values of the binary mass ratio $\qb$: $10^{-3},10^{-2}, 0.1, 0.3, 1$. The lowest value of $\qb$ corresponds to the so-called thermal mass \citep{GR01}, for which $q_b=(H/R)^3$ and below which the disc-perturber coupling is in the linear regime. This allows us to make meaningful comparison with the low-mass planetary case, see Section \ref{sec:comp_prev} for details. At the other end of spectrum, for $\qb$ approaching unity, the disc-binary interaction can be highly non-linear.

After describing the general morphology of the binary-induced perturbation in the disc (Section \ref{sec:res-pert}), we present the results for the $\Texs$ behaviour in Section \ref{sec:res-torque}, followed by the discussion of the nature of torque oscillations in Sections \ref{sec:torque-origin} \& \ref{sec:torque-analysis}.


\subsection{Perturbation to the disc structure}
\label{sec:res-pert}


The top row of Fig.  \ref{fig:polar_sig} shows maps of the surface density of the CBDs at $t=50\Pb$ as they are being perturbed by binaries with four different values of $\qb$. One can see the density waves propagating through the disc from the edge of the central cavity, which become more visible as $\qb$ increases. The cavity itself maintains a roughly axisymmetric shape for $\qb=10^{-2},0.1$, but starts to develop some asymmetries for $\qb=0.3$ at the time shown. In the equal mass case (panel (d)), the cavity is considerably elongated by what appears to be an $m=2$ perturbation --- this shape is different from an eccentric $m=1$ shape often found in CBD simulations at later times.  

The structure of the surface density perturbation $\delta\Sigma$ induced by the binary can be more easily seen in the bottom panels of Fig.  \ref{fig:polar_sig}, where we subtract the axisymmetric component of the surface density $\langle\Sigma\rangle_\phi$ from $\Sigma(R,\phi)$ and plot $\Sigma/\azav{\Sigma} - 1$. This clearly reveals the non-axisymmetric structure of $\delta\Sigma$ for which the binary is responsible. To facilitate the interpretation of these maps, in Fig.  \ref{fig:fourier_dsig} we also plot as a function of $R$ the strength of several low-$m$ Fourier harmonics\footnote{The definition of the Fourier coefficients in this work is a factor of 2 different (greater) from \citet{Cimerman2023b}. } of $\delta\Sigma$ measured in our simulations at a fixed time $t=50\Pb$. 

For low $\qb\leq10^{-2}$ the perturbation has a very regular shape in the form of two tightly-wrapped spiral arms propagating away from the centre. This is the classical density wave with $m=2$ azimuthal periodicity that is launched in the CBD by the binary (see also Section \ref{sec:torque-origin}). Indeed, the number of arms is consistent with Fig.  \ref{fig:fourier_dsig}b, which shows that $m=2$ Fourier component $\delta\Sigma_2$ dominates in the CBD out to $R\lesssim 4.5\ab$ (Fig.  \ref{fig:fourier_dsig}a shows that $\delta\Sigma_2$ is even more prominent for $\qb=10^{-3}$). Moreover, this pattern is stationary in the frame co-rotating with the binary. Assuming a linear perturbation in the form $\delta\Sigma\propto\exp[\I m\left(\phi-\phi_2(t)\right)+\I\int^R k_R(R^\prime)\de R^\prime]$, stationary in the co-rotating frame, the radial wavenumber $k_R$ is given in the WKB limit by
\begin{align}
	k^2_{R} = \cs^{-2}
		\left[ 
			m^2 (\Omega - \OmegaP)^2 - \Omega^2
		\right], 
	\label{eq:k_WKB}
\end{align}
where $\OmegaP$ is the (azimuthal) pattern frequency of the wave mode, which is equal to $\Omegab$ for the binary-driven waves. The phase of the perturbation is preserved along the characteristics given by $\de \phi/\de R=-k_R(R)/m$, i.e. along the curves
\begin{align}
	\phi &=\phi_1(t)+\tilde\phi_m(R),
 \label{eq:crests} \\
 \tilde\phi_m(R)&= \phi_{m,0}-\int^R \frac{\sqrt{\left(\Omega(R^\prime)-\OmegaP\right)^2-m^{-2}\Omega^2(R^\prime)}}{\cs(R^\prime)} \,\de R^\prime,
 \label{eq:phi_m}
\end{align}
where $\phi_{m,0}$ is a constant and we used the relevant root of Eq. (\ref{eq:k_WKB}) for $k_R$. This expression is valid outside the Outer Lindblad Resonance (OLR) at which $k_R=0$; the corresponding radius $R_\mathrm{OLR}$ is such that $\Omega(R_\mathrm{OLR})=\OmegaP \, m/(m+1)$.

We illustrate the WKB prediction (\ref{eq:phi_m}) for the location of the wave-crest of the perturbation with the black dotted curve in Fig.  \ref{fig:polar_sig}e for $m=2$, $\OmegaP=\Omegab$ and by choosing $\phi_{m,0}$ to roughly align with the wave peak in the inner disc. One can see that it traces the shape of the spiral arm very well over large distances in the disc, confirming its identification with the density wave.

According to Fig.  \ref{fig:fourier_dsig}b, for $\qb=10^{-2}$, the $m=1$ component starts to dominate $\delta\Sigma$ outside $4.5\ab$, this can clearly be seen in Fig.  \ref{fig:polar_sig}e right at the outer edge ( $R \simeq 8\ab$). Also, note that at small $R$ each winding of the black dotted WKB curve encompasses two spiral arms, while in the outer disc it encompasses only one. We believe that this transition from $m=2$ to $m=1$ dominance occurs as a result of the spiral arm merger driven by the nonlinear effects: one of the two initial arms has a higher amplitude and propagates somewhat faster \citep{GR01} than the lower-amplitude one, eventually overtaking it and merging with it --- this process is better visualized in Fig.  \ref{fig:dsig_dvort_map}b. To support this interpretation we note that in the lower $\qb=10^{-3}$ simulation such an arm merger occurs much further out in the disc. This is because density wave amplitude is smaller in lower $\qb$ systems (which can be deduced from the vertical scale of the different panels in Fig.  \ref{fig:fourier_dsig}) so that the nonlinear wave evolution is slower as well. 

For higher $\qb=0.1$, the two tightly-wound spiral arms are still clearly visible in the inner disc and merge into a single arm in the outer disc, see Fig.  \ref{fig:polar_sig}f. However, some perturbations to their shape start to emerge. Note that in the near-cavity part of the disc the perturbation map reveals a three-fold symmetry, which does not appear as a strong $m=3$ component in Fig.  \ref{fig:fourier_dsig}c. This is simply because in Fig.  \ref{fig:polar_sig}f the perturbation is normalized by $\azav{\Sigma}$, which becomes very small inside the cavity. The origin of this $m=3$ component will become clear in Section \ref{sec:vortices}.

The trend of increasing irregularity continues at higher $\qb=0.3$, with the tightly-wrapped two-arm spiral arms losing their coherence across the whole domain and a one-sided perturbation emerging at $R\lesssim 4\ab$. Both effects are visible in Fig.  \ref{fig:polar_sig}g, and Fig.  \ref{fig:fourier_dsig}d also reveals a strong $m=1$ component of $\delta\Sigma$ in this radial range. 

\begin{figure*}
\begin{center}
\includegraphics[width=\textwidth]{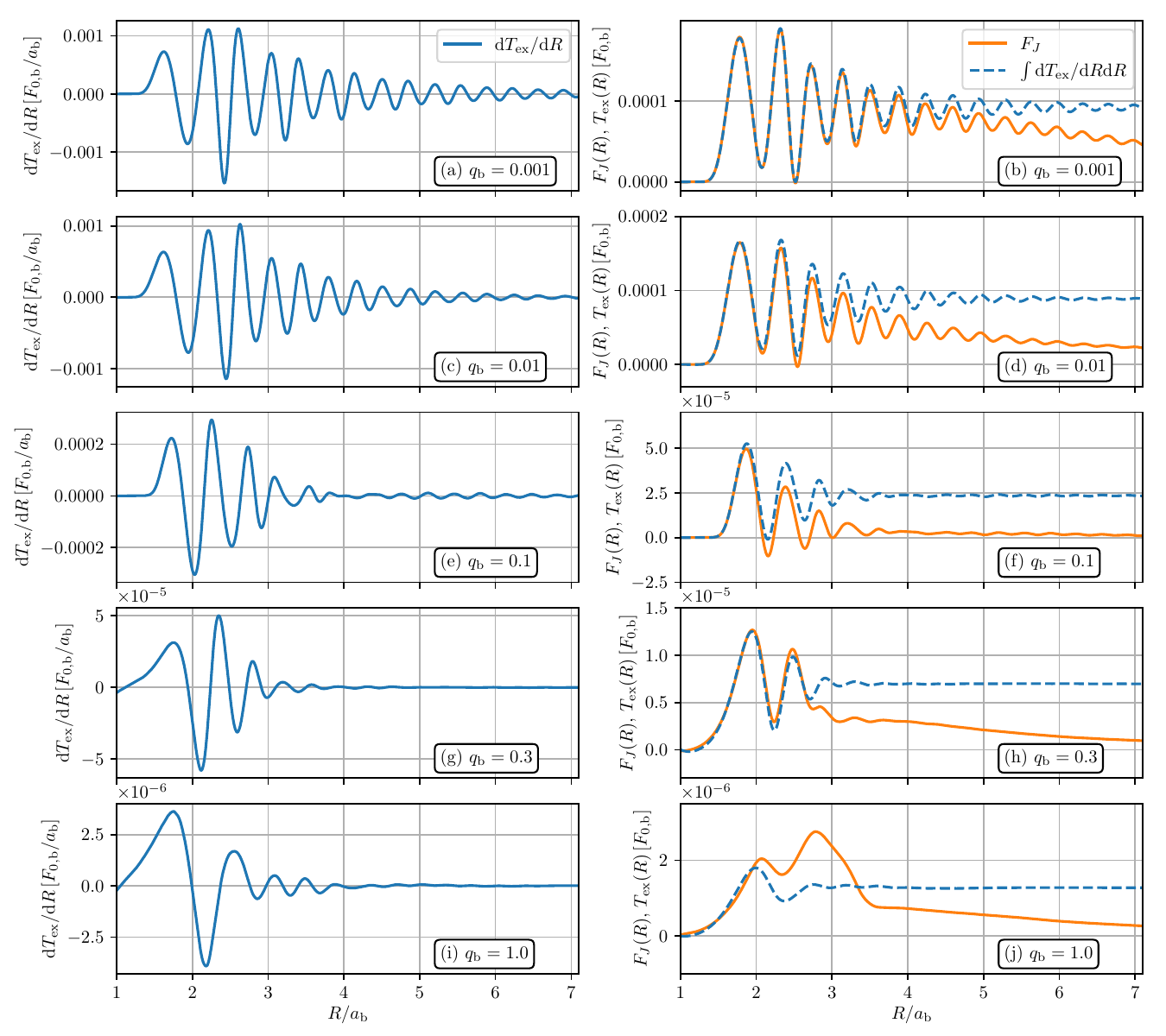}
\caption{{\it Left:} Excitation torque density $\Texs$ in a circumbinary disc extracted from simulations (with $h_0=0.1$) for binaries with different mass ratio $\qb$. 
{\it Right:} Angular momentum flux $F_J(R)$ and cumulative (integrated) torque $T_\mathrm{ex}(R)$.
All quantities are averaged over 10 orbits from $t = 45 \Pb$ to $t = 55 \Pb$ using 100 samples per orbit. Total torque and $F_J$ are expressed in units of $F_{J,\mathrm{b}} \propto \qb^2$ given by Eq. (\ref{eq:FJ_b}). See Section \ref{sec:res-torque} for details.
}
\label{fig:binary_dTdr_N}
\end{center}
\end{figure*}

In the equal-mass case ($\qb=1$) the perturbation pattern looks quite different from the low-$\qb$ runs. Figure \ref{fig:polar_sig}h shows that while $\delta\Sigma_2$ is still the dominant component of the perturbation (see Fig.  \ref{fig:fourier_dsig}e), it is no longer tightly-wrapped as in e.g. the $\qb=10^{-2}$ case. What we see instead are the strong, open $m=2$ spiral arms that are also clearly perturbed by a more-tightly wound pattern (sharp 'feathers' emanating from the prominent open spiral arms).

An explanation of these trends in the behaviour of $\delta\Sigma(R,\phi)$ as a function of $\qb$ --- a gradual transition from tightly-wrapped to more open spiral arms and the general increase of irregularity of $\delta\Sigma$ with growing $\qb$ --- will be provided in Section \ref{sec:vortices}.

\begin{figure*}
	\begin{center}
	\includegraphics[width=\textwidth]{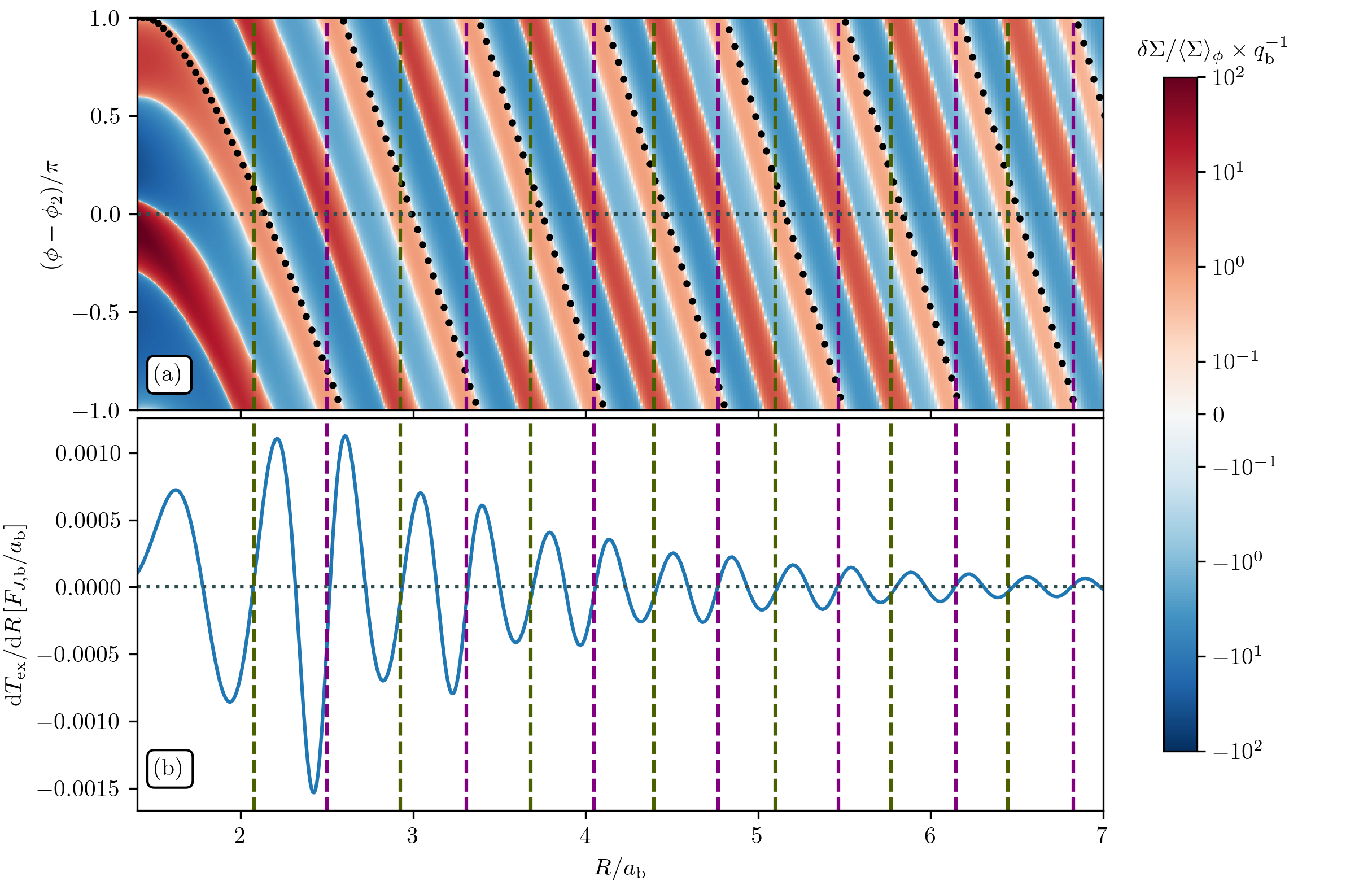}
	\caption{
        Illustration of the origin of $\Texs$ oscillations using a globally isothermal simulation with $h_0=0.1$ and $\qb=10^{-3}$: (a) a map of $\Sigma/\azav{\Sigma}-1$ illustrating the two-armed shape of the density waves in the CBD in Cartesian $R-\phi$ projection, (b) $\Texs$ plotted on the same radial scale. Black dotted curve in panel (a) shows the linear WKB prediction (\ref{eq:phi_m}) for the wave arm shape (with $m=2$, $\OmegaP=\Omegab$ and some chosen $\phi_{m,0}$), which provides an excellent fit. Vertical violet and green markers show the radii where the peaks of each spiral arm align with the binary such that $\phi=\phi_2$. These markers clearly show that $\Texs$ periodicity is directly related to the way the density waves wrap around the origin. 
	}
	\label{fig:dtdr_qem3_markers}
	\end{center}
\end{figure*}

\begin{figure}
	\begin{center}
	\includegraphics[width=0.49\textwidth]{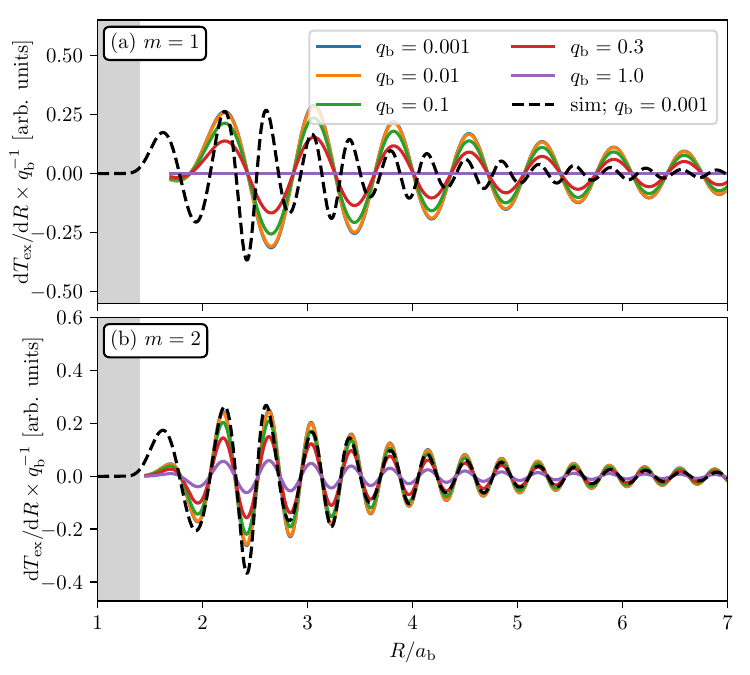}
	\caption{
		Torque density (re-scaled by the binary mass ratio $\qb$) due to an artificially imposed surface density perturbation that assumes $\delta \Sigma/\Sigma(R,t=0) = $ const and has only a single azimuthal Fourier component in the form of spiral arm(s) co-rotating with the binary, with the shape given by the linear WKB relation (\ref{eq:phi_m}): $\delta \Sigma \propto \Sigma(R,t=0) \exp \lbrace \I m [\phi - \phi_2(t)-\tilde\phi_m(R)] \rbrace$.
		The top panel shows $m=1$, the bottom $m=2$, for a range of $\qb$ (colours). The initial phase $\phi_{m,0}$ in $\tilde\phi_m(R)$ is adjusted to match the location of the first peak of the $\Texs$ obtained from a $\qb=10^{-3}$ simulation (black dashed curve, arbitrarily rescaled for presentation purposes to match the first significant peak of $\qb^{-1}\Texs$ in amplitude). The grey shaded area indicates the heavily depleted cavity region. The fact that this toy model reproduces $\Texs$ periodicity well for $m=2$ and all $\qb$ means that it is the two-armed spiral with the pattern speed $\OmegaP = \Omegab$ that sets the periodicity of torque oscillations in our $\qb=10^{-3}$ simulation.
	}
	\label{fig:toy_comb}
	\end{center}
\end{figure}


\subsection{Torque Density Profiles}
\label{sec:res-torque}


Next we discuss our results on the torques characterizing disc-binary coupling. In the left column of Fig. \ref{fig:binary_dTdr_N}, we show the radial profiles of $\Texs$ for five different values of $\qb$ that we consider, assuming the CBD setup with the initial $\Sigma(R)$ distribution in the form (\ref{eq:Sigma0}). By examining these profiles, we can make the following observations.

First, $\Texs$ exhibits oscillatory behaviour with $R$ for all $\qb$ that we examined. Moreover, oscillations are clearly quasi-periodic, i.e. the radial spacing of the locations where $\Texs=0$ follows a particular pattern, at least approximately. Second, the amplitude of $\Texs$ oscillations, when normalized by $F_{J,\mathrm{b}}$, decreases with $\qb$. However, in the lowest-$\qb$ runs, $\qb=10^{-3}$ and $10^{-2}$, the (scaled) amplitude is almost the same, and there is an overall similarity of the scaled $\Texs$ profiles. Third, the amplitude of $\Texs$ oscillations decays with $R$ either from the start (for high $\qb$) or beyond $R\approx 2.5\ab$ (for $\qb\lesssim 0.1$). This decay with $R$ is faster for higher $\qb$. Fourth, the radial period of $\Texs$ oscillations is somewhat larger for higher $\qb$ closer to the cavity (even though there are fewer clear oscillation cycles in these runs). These trends in the behaviour of $\Texs$ with $R$ and $\qb$ will be explained in Sections \ref{sec:torque-origin} \& \ref{sec:torque-analysis}.

In the right column of Fig. \ref{fig:binary_dTdr_N} we show other important angular momentum metrics: the integrated gravitational torque $T_\mathrm{ex}(R)=\int_0^R\left(\de T_\mathrm{ex}(R^\prime)/\de R\right) \de R^\prime$ (blue dashed curve), and angular momentum flux $F_J$ (orange curve) defined by Eq. (\ref{eq:FJ}). The integrated torque $T_\mathrm{ex}(R)$ is a measure of the angular momentum injected (up to radius $R$) by the non-axisymmetric component of the binary gravity into the density waves propagating through the disc. Angular momentum flux $F_J$ is the amount of angular momentum that these density waves carry across radius $R$ per unit of time. Thus, the two metrics are related (see below).

One can see that $T_\mathrm{ex}(R)$ saturates at a finite and positive value for all $\qb$. This implies that gravitational torque due to the binary endows density waves with a {\it positive} angular momentum, as expected \citep{Meyer1987}. At the same time, $F_J$ decays as $R$ increases. This happens since the density waves not only accumulate angular momentum due to the binary torque but also lose it as they dissipate due to the their non-linearity \citep{GR01,R02}. As a result, while initially $F_J$ tends to replicate the oscillatory features of $T_\mathrm{ex}(R)$ (especially at low $\qb$), eventually the two curves diverge from each other and $F_J$ falls below $T_\mathrm{ex}(R)$. This divergence is faster for higher $\qb$ because the density waves become more nonlinear as $\qb$ increases, as we noted in Section \ref{sec:res-pert}. 

If the binary torque were the only driver of the perturbation in the disc, the difference between $T_\mathrm{ex}(R)$ and $F_J$ would constitute the deposition torque density $\Tdeps$ due to the binary (defined in Section \ref{sec:intro}), which characterizes the amount of angular momentum passed by the density wave from the binary to the disc fluid  \citep{R02b,Dong2011,Cimerman2021,Dempsey2020}. In the CBD context, $\Tdeps$ should be positive as the gravitational torque of the binary drives positive angular momentum flux through the disc. However, this is not the case in Fig.  \ref{fig:binary_dTdr_N}j for $\qb=1$, which  shows that $F_J(R)$ {\it exceeds} $T_\mathrm{ex}(R)$ over a radial range $(2-3.3)\ab$; in fact,  a slight increase of $F_J(R)$ above $T_\mathrm{ex}(R)$ near the first peak can be seen also for $\qb=0.3$ in panel (i) of that figure. Interestingly, for these values of $\qb$ one also finds $F_J$ not decaying to zero rapidly, as for lower $\qb$; instead, it falls off with $R$ rather slowly. These intriguing anomalies will be explained in Section \ref{sec:vortices}.


\subsection{Origin of the Torque Density Oscillations}
\label{sec:torque-origin}


Figure \ref{fig:dtdr_qem3_markers} provides a graphic explanation for the origin of the torque oscillations, using the results of the $\qb=10^{-3}$ simulation as a basis. In panel (a) we plot the relative non-axisymmetric perturbation amplitude $\Sigma/\azav{\Sigma} - 1$, similar to Fig.  \ref{fig:polar_sig}e, but now in Cartesian $(R,\phi)$ coordinates. The black dotted curve again shows the linear WKB prediction (\ref{eq:crests})-(\ref{eq:phi_m}) for shape of the wake, which matches the shape of the two spiral arms very well. We have chosen a low value of $\qb$ to illustrate our point simply because we want to minimize the effects of wave nonlinearity, which would otherwise complicate our interpretation. 

In panel (b) we reproduce $\Texs$ profile (as in Figs.  \ref{fig:Airy}, \ref{fig:binary_dTdr_N}a) on the same radial scale as in panel (a). Vertical violet and green dashed lines running through both panels mark the radial locations where the peak of each spiral arm in panel (a) crosses the line\footnote{The periodicity of $\Texs$ can be demonstrated by using any other value of $\phi$ as well.} $\phi=\phi_2$ running from $M_1$ to $M_2$. One can see that these lines cross $\Texs$ curve near its every second node, matching the radial periodicity of $\Texs$ remarkably well. Thus, this figure\footnote{Figures \ref{fig:dtdr_qem3_markers} \& \ref{fig:toy_comb} are similar to Figs. 3 \& 4 of \citet{Cimerman2023b} illustrating the origin of the torque wiggles, which are later discussed in Section \ref{sec:planet}.} clearly shows that the periodicity of $\Texs$ in $R$ is set entirely by the shape of the two binary-driven spiral arms as they wrap around the origin. In fact, this connection can be gleaned already from \citet{MacFadyen2008} by noting that the locations of $\Texs$ peaks in their Fig. 4 exhibit the same radial periodicity as the locations,  at which the two-armed spiral wake crosses the line $\phi=0$ in their Fig. 5.

Figure \ref{fig:toy_comb} takes this point even further by showing $\Texs$ (re-scaled by $\qb$) in a CBD with an artificially imposed perturbation pattern in the form $\delta \Sigma \propto \Sigma(R,t=0)\exp \lbrace \I m [\phi - \phi_2(t)-\tilde\phi_m(R)] \rbrace$, i.e. a single $m$-th azimuthal harmonic of constant amplitude relative to $\Sigma(R,t=0)$ given by Eq. (\ref{eq:Sigma0}), which forms spiral arm(s) with the shape given by the linear WKB equation (\ref{eq:phi_m}) in the frame co-rotating with the binary, very similar to the arms in Fig.  \ref{fig:dtdr_qem3_markers}a. This calculation is done for $m=1,2$ and several values of $\qb$, with the true $\Texs$ from a $\qb=10^{-3}$ simulation plotted for illustration as a black dashed line 

One can see that $m=2$ calculation in panel (b) reproduces the periodicity of the true $\Texs$ extremely well, independent of $\qb$ --- the value of $\qb$ affects only the amplitude of torque oscillations (in this exercise the amplitude of the imposed perturbation is independent of $\qb$). At the same time, for $m=1$ in panel (a) the periodicity has twice the radial length scale of the $\Texs$ from simulations. This is consistent with the idea that it is the two-armed spiral in our simulation with the shape closely following Eq. (\ref{eq:phi_m}) that sets the periodicity of $\Texs$. This toy model also makes it obvious that the radial periodicity of $\Texs$ is directly related to the azimuthal wavenumber of the dominant Fourier harmonic of the density wave perturbation. These findings will be explained further in Section \ref{sec:torque-analysis}

In conclusion we also point out that in Fig.  \ref{fig:dtdr_qem3_markers}, the green markers, associated with the stronger spiral arm, steadily move closer to the violet ones (for the weaker arm) as $R$ increases. This is yet another illustration of the nonlinear wave propagation (the front of the stronger shock travels faster catching up with the weaker one) which will ultimately result in the merger of these arms into one. 



\subsection{Theoretical analysis}
\label{sec:torque-analysis}


Next, we provide a more quantitative analysis of the $\Texs$ oscillations in CBDs. In Appendix \ref{app:derive}, we demonstrate that the definition (\ref{eq:Texs}) for $\Texs$ can be re-cast in the following form:
\begin{align}
\Tex &= G \Mt \sum\limits_{n=1}^\infty n\, C_n(R)\, A_n(R),
\label{eq:Texs3}\\
C_n(R) &=\int\limits_0^{2\pi}\Sigma(R,\phi)\sin[n(\phi-\phi_2(t))]\,\de \phi,
\label{eq:Cn}\\
A_n(R) &= (1+\qb)^{-1} \left[(-1)^n b^{(n)}_{1/2} (\alpha_1)+\qb  b^{(n)}_{1/2} (\alpha_2)\right],
\label{eq:An}
\end{align}
where $\phi_2(t)$ is the azimuthal angle towards $m_2$, $\alpha_i=R_i/R$, $i=1,2$,  and $b^{(n)}_{1/2}(\alpha)$ are the Laplace coefficients defined by Eq. (\ref{eq:lap_def}). The functions $C_n(R)$ absorb all information about the spatial structure of $\Sigma$ in the disc. Since for all $n\ge1$ the axisymmetric component of $\Sigma(R,\phi)$ does not contribute to $C_n$, we can replace $\Sigma(R,\phi)$ in Eq. (\ref{eq:Cn}) with $\delta\Sigma(R,\phi)$ --- the perturbation of the surface density in the CBD.

Note that according to Eq. (\ref{eq:Cn}), the time-averaged torque density $\langle\Texs\rangle_t$ (which is what we display in Fig.  \ref{fig:binary_dTdr_N}) is determined only by the component of $\delta\Sigma$, which is stationary in the frame co-rotating with the binary. In other words, only $\delta\Sigma(R,\phi)$ in the form $\delta\Sigma(R,\phi-\phi_2(t))$ can give rise to non-zero $C_n$, while all other components of $\delta\Sigma(R,\phi)$ varying in the co-rotating frame would result in no net $\langle\Texs\rangle_t$, given a long enough averaging period. This point will be discussed further in Section \ref{sec:vortices} (see also Fig.  \ref{fig:time-average}).

Motivated by this observation, we now consider the following WKB-inspired azimuthal Fourier decomposition of $\delta\Sigma(R,\phi)$, which is stationary in the frame co-rotating with the binary:
\begin{align}
\delta\Sigma(R,\phi)=\sum\limits_{m=1}^\infty |\delta\Sigma_m(R)|\cos\left[m\left(\phi-\phi_2(t)-\tilde\phi_m(R)\right)\right],
\label{eq:dSig}
\end{align}
where $\tilde\phi_m(R)$ is given by Eq. (\ref{eq:phi_m}), and $|\delta\Sigma_m(R)|$ is the amplitude of the $m$-th harmonic propagating according to its WKB dispersion relation. Note that the Fourier components shown in Fig.  \ref{fig:fourier_dsig} are based on instantaneous $\delta\Sigma$ and are somewhat different (especially for high $\qb$) from $\delta\Sigma_m$ computed from $\delta\Sigma$ time-averaged in the binary frame, see Section \ref{sec:vortices}.

Plugging the ansatz (\ref{eq:dSig}) into Eq. (\ref{eq:Cn}) and integrating, we can re-write (\ref{eq:Texs3})-(\ref{eq:An}) as
\begin{align}
\Tex &= \sum\limits_{m=1}^\infty \Texm,
\label{eq:Texs4}\\
\Texm &= \pi \, G  \Mt \, m  \, |\delta\Sigma_m(R)| \, A_m(R) \,  \sin \left(m\tilde\phi_m(R)\right).
\label{eq:Texm}
\end{align}
This expression provides a mathematical basis for understanding the nature of $\Texs$ oscillations.  We verified that for $\delta\Sigma_m/\Sigma(R,t=0)=\const$, Eq. (\ref{eq:Texm}) reproduces exactly $\Texs=\Texms$, for $m=1$ and $2$ shown by the coloured curves in Fig. \ref{fig:toy_comb}a,b, correspondingly. 

As illustrated by Fig. \ref{fig:fourier_dsig}, in CBDs the perturbation $\delta\Sigma$ is typically dominated by one or two azimuthal harmonics $\delta\Sigma_m$. The most common situation is the dominance of $m=2$ pattern (e.g. for $\qb=10^{-3},10^{-2}$ or $1$; see also Fig. \ref{fig:time-average}), although $m=1$ or $3$ harmonics may also play a role, at least over a range of radii (e.g. for $\qb=10^{-2},0.1$).  Assuming that $\delta\Sigma$ is indeed dominated by a single $m$-th harmonic,  such that $\Texs\approx \Texms$, we can provide a mathematical justification for several observations made in Section \ref{sec:res-torque}.

\begin{figure}
	\begin{center}
	\includegraphics[width=0.49\textwidth]{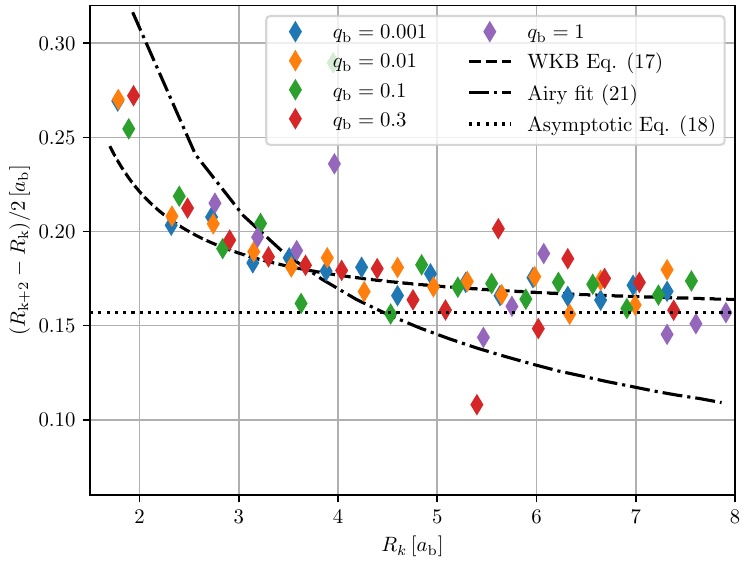}
	\caption{
		Radial spacing of the nodes of $\Texs$ --- the distance between the consecutive radial locations $R_k$ where $\Texs$ becomes zero (approximated as $(R_{k+2}-R_k)/2$ for symmetry) --- plotted as a function of $R_k$. Points of different colour correspond to our fiducial inviscid simulations (globally isothermal with $h_0=0.1$) with different $\qb$. Black dashed line shows $R_{k+1}-R_k$ computed as a continuous function of $R_k$ using the analytical WKB relation (\ref{eq:dR_k}) with $m=2$, providing a good fit to the simulation data (coloured points). The dotted line reflects the asymptotic relation (\ref{eq:dR_k1}). The dot-dashed curve represents the asymptotic 'Airy' fit (\ref{eq:dR_k-res}) discussed in Section \ref{sec:resonant-torque}.
	}
	\label{fig:spacing_zeros_dtdr_var_q}
	\end{center}
\end{figure}

First, the oscillatory behaviour of $\Texms$ with $R$ immediately follows from Eq. (\ref{eq:Texm}) through the factor $\sin \left(m\tilde\phi_m(R)\right)$. This dependence also naturally explains the quasi-periodicity of the radial oscillations of $\Texs$. Indeed, it predicts the nulls of $\Texms$ to occur at $R_k$, $k=1,..$ given by the condition $\tilde\phi_m(R_k)=k\pi/m$. In particular, Eq. (\ref{eq:phi_m}) with $\OmegaP=\Omegab$ gives the following WKB relation between the consecutive nodes $R_k$ and $R_{k+1}$ of $\Texs$
\begin{align}
 \int^{R_{k+1}}_{R_k} \cs^{-1}(R^\prime)\sqrt{\left(\Omega(R^\prime)-\Omegab\right)^2-m^{-2}\Omega^2(R^\prime)} \,\de R^\prime=\frac{\pi}{m}.
 \label{eq:dR_k}
\end{align}
Outside the cavity, where $\Omega\ll \Omegab$, the distance between the consecutive nodes can be approximated in the linear WKB limit as
\begin{align}
 R_{k+1}-R_k \approx \frac{\pi}{m}\frac{\cs(R_k)}{\Omegab},
 \label{eq:dR_k1}
\end{align}
explaining the regular nature of the $\Texs$ periodicity with $R$ (this lengthscale is constant for $\cs=\const$). The direct dependence of the radial length scale of the torque oscillations on $m$ has been previously illustrated by a toy model in Fig. \ref{fig:toy_comb}. 

\begin{figure*}
	\begin{center}
	\includegraphics[width=\textwidth]{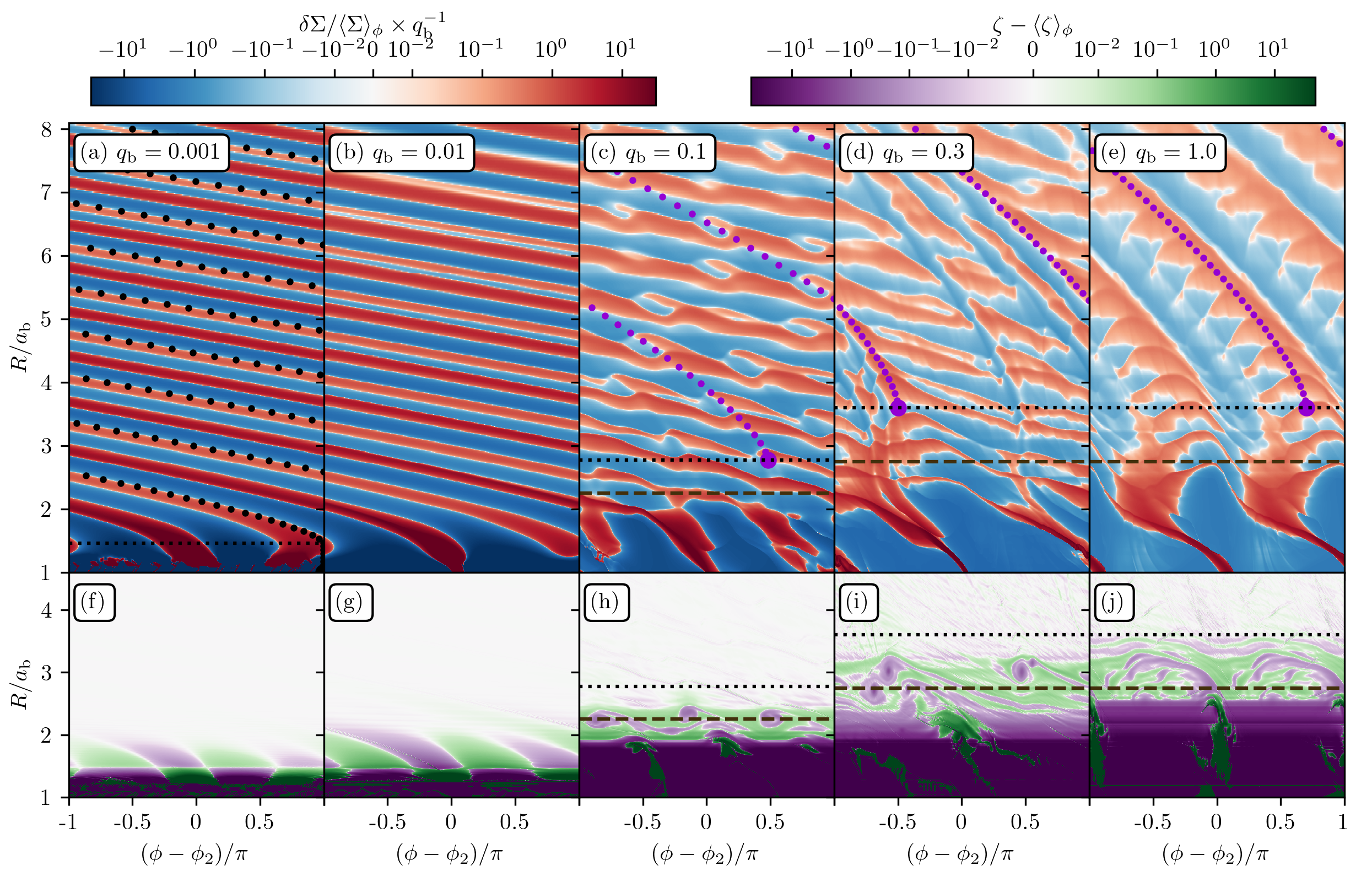}
	\caption{Snapshots at $t = 50\Pb$ of (top row) relative surface density perturbation $\delta\Sigma/\azav{\Sigma}$, scaled by $\qb$ and (bottom row) vortensity deviation $\zeta-\azav{\zeta}$ from its azimuthal average $\azav{\zeta}$, plotted for different $\qb$. Black dotted curve in panel (a) shows the linear WKB prediction (\ref{eq:phi_m}) for the wave arm shape for $m=2$ and $\OmegaP=\Omegab$ (with some chosen $\phi_{2,0}$) providing an excellent match to the shape of the low-amplitude spiral arms driven by the $\qb=10^{-3}$ binary. In panels (c)-(e) we show WKB fits (\ref{eq:phi_m}), (\ref{eq:pitch}) for vortex-driven density waves (violet dots), in which we use $(m,\OmegaP)=(3, 0.33\Omegab),
	(2, 0.22\Omegab),(2, 0.23\Omegab)$ for $\qb = 0.1,0.3,1$, respectively. Horizontal black dotted lines mark (in both rows) the outer Lindblad resonance radii for these $(m,\OmegaP)$. The dark green dashed lines mark the corotation radii of these modes (i.e. radii where $\azav{\Omega(R)} = \OmegaP$), and they fit well the radial locations of the cores of the vortices clearly visible in the panels (h)-(j). Going from left to right, this plot shows the transition from the binary-driven spirals dominating to the vortex-driven modes dominating. 
	}
	\label{fig:dsig_dvort_map}
	\end{center}
\end{figure*}

In Fig. \ref{fig:spacing_zeros_dtdr_var_q} we illustrate the relation (\ref{eq:dR_k}), assuming that the binary drives a two-armed ($m=2$) spiral pattern, by plotting it as a continuous function of $R_k$ (and approximating $R_{k+1}-R_k\approx (R_{k+1}-R_k)/2$) in the globally isothermal case with $h_0=0.1$; the horizontal dotted line shows the asymptotic limit (\ref{eq:dR_k1}). We also plot (as diamonds) the data for $(R_{k+2}-R_k)/2$ taken from our fiducial simulations with different values of $\qb$. One can see that the WKB prediction (\ref{eq:dR_k}) provides a good fit to the clustering of the simulation data points, supporting our interpretation of the torque oscillations put forward in Section \ref{sec:torque-origin}. In fact, given the diverse morphologies of $\delta\Sigma$ in panels (e)-(h) of Fig. \ref{fig:polar_sig}, it may seem surprising that the agreement between the relation (\ref{eq:dR_k}) for $m=2$ and simulation data for very different $\qb$ is so good; this agreement will be explained later in Section \ref{sec:vortices}. The scatter of data points around the relation (\ref{eq:dR_k}) is caused by the fact that other Fourier components of $\delta\Sigma$ (i.e. $m\neq 2$) are also contributing to $\Texs$ at some level in the simulations (also, for higher $\qb$ we get only a few clearly pronounced oscillations in the inner disc). 

Second, the decrease of the amplitude of $\Texs$ normalized by $F_{J,\mathrm{b}}$ with increasing $\qb$ is due to the nonlinear effects. In the linear limit and, very importantly, in a disc with a fixed surface density, the normalization by $F_{J,\mathrm{b}}$ makes the disc-perturber coupling independent of the perturber mass, see \citet{GT80} \& \citet{Cimerman2021}. This is roughly the case for the lowest\footnote{Strictly speaking, the linear approximation applies for $\qb\lesssim (H/r)^3\approx 10^{-3}$.} $\qb=10^{-3},10^{-2}$, which look strongly alike, see Fig. \ref{fig:binary_dTdr_N}a,b. However, for higher $\qb$ the cavity size starts to change ($\Rcav$ increases with $\qb$, see Fig. \ref{fig:sig_r}), reducing the gas surface density near the important resonances where the density wave is launched. As a result, the relative amplitude of the torque oscillations (normalized by $F_{J,\mathrm{b}}$) decreases as $\qb$ increases. 

Third, according to Eq. (\ref{eq:Texm}) the evolution of the amplitude of $\Texs$ oscillations with $R$ is determined by the behaviour of both the mode amplitude $|\delta\Sigma_m(R)|$ and the scaling factor $A_m(R)$, which characterizes the weakening of the binary potential with the distance. Figure \ref{fig:fourier_dsig} shows that $|\delta\Sigma_m(R)|$ initially increases with $R$ due to the initial wave excitation but then decays as a result of non-linear dissipation, which in the absence of explicit linear dissipation \citep[e.g.][]{Miranda2020I} is the only wave damping mechanism available \citep{GR01}. This figure also shows that $|\delta\Sigma_m(R)|$ decay is faster for higher $\qb$ as a result of stronger wave nonlinearity. For that reason the amplitude of $\Texs$ oscillations decreases over shorter radial interval in high $\qb$ cases. But even if $|\delta\Sigma_m(R)|$ were to stay unchanged, the amplitude of $\Texs$ oscillations would still go down as $R\to \infty$ due to the $A_m(R)$ behaviour alone, since $b_{1/2}^{(m)}(\alpha)\propto \alpha^m$ as $\alpha\to 0$. 

Fourth, wave nonlinearity also explains why the radial period of $\Texs$ oscillations near the cavity edge tends to slightly increase with $\qb$. Figure \ref{fig:spacing_zeros_dtdr_var_q} provides some evidence for this, as the higher-$\qb$ points tend to lay above the linear relation (\ref{eq:dR_k}) for $R\lesssim 4\ab$, where the wave nonlinearity is strongest. Indeed, it is known from the studies of disc-planet interaction \citep{R02,Zhu2015,Cimerman2021} that the nonlinear spiral density waves launched by massive planets have higher opening angles than the wave launched by the low-mass planets. For the same reason, in our case the more nonlinear density waves in higher $\qb$ runs travel further in $R$ per each azimuthal wrapping than in the low-$\qb$ cases, explaining the (weak) evolution of the radial period of torque oscillations with $\qb$. Note that this effect is not captured in equations (\ref{eq:dR_k})-(\ref{eq:dR_k1}) which are based on the linear approximation (\ref{eq:crests})-(\ref{eq:phi_m}) for the global shape of the density wave.  


\section{Vortex-driven waves }
\label{sec:vortices}


In Sections \ref{sec:res-pert} and \ref{sec:res-torque} we pointed out several interesting anomalies in CBD behaviour as $\qb$ approaches unity: the emergence of very open spiral arms and increased irregularity of the perturbation pattern in Fig. \ref{fig:polar_sig}h, as well as the unexpected excess of $F_J(R)$ over $T_\mathrm{ex}(R)$ in a finite radial range in Fig. \ref{fig:binary_dTdr_N}h,j. To further investigate these anomalies we examined the spatial structure of the fluid vortensity $\zeta=\Sigma^{-1}\left(\nabla\times {\bf u}\right)$, which provides a useful diagnostic of various processes in discs \citep{Coleman2022a,Cimerman2021}. 

Figure \ref{fig:dsig_dvort_map} illustrates our findings for different values of $\qb$ at $t=50\Pb$. The lower panels of this figure show the non-axisymmetric component of the vortensity $\zeta-\langle\zeta\rangle_\phi$ in the inner part of the CBD. The use of the non-axisymmetric component allows us to remove the dominant axisymmetric vortensity contribution and focus on any structures that may be forming near the cavity edge. The top panels of Fig. \ref{fig:dsig_dvort_map} show contemporaneous snapshots of the relative surface density perturbation (normalized by $\qb$) in a Cartesian $(R,\phi)$ projection (i.e. as in Fig. \ref{fig:dtdr_qem3_markers}a). The use of this projection helps us visualize the pitch angle $\theta$ of the spiral features (the angle between the tangent to the spiral and the local azimuthal direction) in the CBD, which is given by
\begin{align}
\tan\theta	=\left|\frac{1}{R}\frac{\de R}{\de \phi}\right| = \frac{m}{R k_R(R)}\approx \frac{\cs(R)}{R\OmegaP},
\label{eq:pitch}
\end{align}
such that the slope of the wave crests in a Cartesian map is $R\tan\theta$; in the last approximation in (\ref{eq:pitch}) we used the WKB expression (\ref{eq:k_WKB}) for $k_R$ and assumed $R\gtrsim R_\mathrm{OLR}$, $\Omega\ll\OmegaP$.

The black dotted curve in panel (a) again illustrates the WKB prediction (\ref{eq:crests})-(\ref{eq:phi_m}) for the shape of the binary-driven spiral in a $\qb=10^{-3}$ CBD assuming $m=2$ and $\OmegaP=\Omegab$. As was already evident in Figs.  \ref{fig:polar_sig}e and \ref{fig:dtdr_qem3_markers}a, it fits the shape of the spiral arms very well. The slow evolution of the two arms towards their merger into one as $R$ increases (see Section \ref{sec:res-pert}) can also be traced in this panel. This picture hardly changes for $\qb=10^{-2}$ (panel (b)), only the merger of the two arms into one is more obvious. The vortensity maps for these values of $\qb$ (panels (f), (g)) show some non-trivial activity only inside the cavity but not in the bulk of the CBD.

Things change noticeably for $\qb=0.1$ in panel (c) as the $\delta\Sigma$ map starts showing irregularities: while the tightly wound spiral arms are still visible in this map, their crests (still well described by the WKB prediction) are clearly perturbed by some additional quasi-regular pattern. For $\qb=0.3$ (panel (d)) this additional pattern becomes more pronounced in the form of a number of open spiral arms that compete in strength with the underlying original tightly wrapped spirals. Finally, in panel (e) (for $\qb=1$), $\delta\Sigma$ is clearly dominated by the two open spiral arms perturbed by the remnants of the original tightly-wrapped spirals, which have now become subdominant. These open spiral arms are clearly seen in Fig. \ref{fig:polar_sig}d,h.

Vortensity maps in panels (h)-(j) reveal the reason for these changes in the appearance of $\delta\Sigma$. For $\qb=0.1$ one can see three highly localized (in $R$ and $\phi$) vortensity concentrations at $R\approx 2.2\ab$, at the 'wall' of the cavity, see Fig. \ref{fig:sig_r}. These concentrations represent fluid vortices, roughly equally spaced in $\phi$ and forming $m=3$ vortensity structure, explaining the three-fold symmetry of $\delta\Sigma$ perturbation inside the cavity in Fig. \ref{fig:polar_sig}f. For $\qb=0.3$ one can again see three sharp vortices, although two of them are in the process of merging with each other at this time (at $\phi-\phi_\mathrm{b}\approx -0.6\pi$). They are also located at larger radii, around $3\ab$, near the peak of $\Sigma$ in Fig. \ref{fig:sig_r}. Finally, for $\qb=1$ one can see two big vortical 'rolls' around $R\approx 3\ab$, which are quite extended in the azimuthal direction and form an $m=2$ pattern.

These different vortical structures (we will generally refer to them as just 'vortices') are the reason behind the emergence of additional open spiral arms in panels (c)-(e). Indeed, localized concentrations of vortensity are known to launch density waves in discs \citep{Li2001,Johnson2005,Mamat2007,Heinemann2009}, similar to embedded planets, and this is what happens in the present case. The vortices visible in panels (h)-(j) drive spiral density waves in the CBD --- the vortex-driven modes discussed in \citet{Coleman2022a} in the context of astrophysical boundary layers. These vortices are passively advected with the disc flow and orbit at the angular frequency $\Omega_\mathrm{vrt}$ corresponding to the radii $R_\mathrm{vrt}$ at which their peaks are located. Since $R_\mathrm{vrt}>\ab$, one naturally has  $\Omega_\mathrm{vrt}<\Omegab$. According to Eq. (\ref{eq:pitch}), this makes the vortex-driven spiral arms (with $\OmegaP=\Omega_\mathrm{vrt}$) more open, i.e. have higher $\theta$, than the tightly-wrapped binary-driven spirals (for which $\OmegaP=\Omegab$).

To confirm that the open spiral arms appearing for $\qb\ge 0.1$ are indeed the vortex-driven density waves, in panels (c)-(e) we draw (violet dotted curves) the WKB prediction for the shape of the vortex-driven modes. We do this using Eqs. (\ref{eq:crests})-(\ref{eq:phi_m}) and $\OmegaP=\Omega_\mathrm{vrt}$, $m$ determined based on vortex characteristics from panels (h)-(j), see figure caption. We show the corotation radii of the modes, coinciding with $R_\mathrm{vrt}$, with dashed dark green lines and the corresponding outer Lindblad resonance locations $R_\mathrm{OLR}$ with horizontal black dotted lines. One can see that the shape of the open vortex-driven spirals is matched by the WKB density wave predictions very well, solidifying their interpretation as vortex-driven modes.

\begin{figure}
	\begin{center}
	\includegraphics[width=0.49\textwidth]{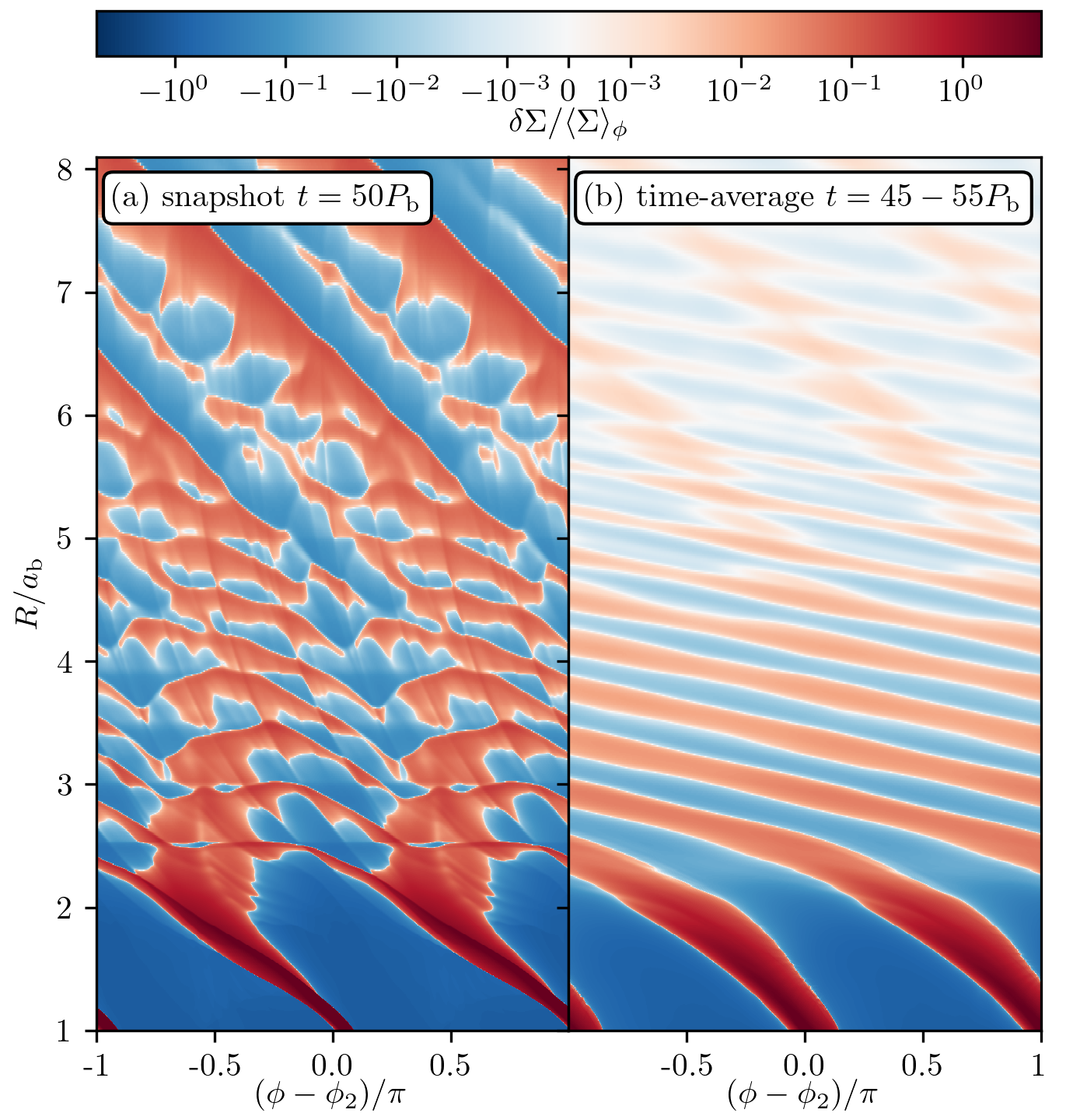}
	\caption{
		Comparison of (a) a surface density perturbation snapshot at $t=50\Pb$ and (b) time-average of the same variable in the frame co-rotating with the binary over the interval $(45-55)\Pb$. Data are from the globally isothermal simulation with $h_0=0.1$ and $\qb=1$. One can see that time-averaging in the frame co-rotating with the binary effectively eliminates the vortex-driven perturbation in the inner disc.
 }
	\label{fig:time-average}
	\end{center}
\end{figure}

We can now address the origin of the anomalies in the behaviour of $F_J$ and $T_\mathrm{ex}$ visible in Fig. \ref{fig:binary_dTdr_N}j, namely the excess of $F_J(R)$ over $T_\mathrm{ex}(R)$ over some radial ranges. If the binary were the only actor responsible for the wave activity in the CBD, one would always expect $F_J(R)<T_\mathrm{ex}(R)$, which is the case in other panels in that figure. However, strong vortices present near the cavity edge change this picture since the density waves that they launch also transport angular momentum through the disc \citep{Paardekooper2010,Coleman2022b}. Since density wave launching by vortices does not involve gravity \citep{Mamat2007,Heinemann2009}, there is no direct contribution from these modes to the gravitational $\Texs$. However, the vortex-driven modes do carry positive angular momentum flux outside $R_\mathrm{vrt}$, providing an additional positive contribution to the $F_J$ carried by the (tightly-wrapped) binary-driven spirals, such that the combined $F_J$ ends up exceeding $T_\mathrm{ex}$. 

\citet{Coleman2022b} also found that the vortex-driven modes excited in the boundary layers of accretion discs decay with the distance due to non-linear dissipation rather weakly, leading to a slow decay of their $F_J$ (see figs. 2, 3 of that paper). We observe a similar phenomenon in panels (i),(j) of the Fig. \ref{fig:binary_dTdr_N}, in which $F_J$ decays rather slowly for $R\gtrsim 3\ab$ --- for these two $\qb$ the vortex-driven modes dominate in the outer disc.

One may wonder why the strong, open vortex-driven spiral arms, asymmetric in the binary frame, do not provide a distinct contribution to $\Texs$ in Fig. \ref{fig:binary_dTdr_N} for $\qb\ge 0.3$. Indeed, Eqs. (\ref{eq:Texs4})-(\ref{eq:Texm}) would seem to predict such a contribution (scaling {\it linearly} with $\qb$, unlike the regular $\Texs$) with $\tilde\phi_m(R)$ determined by the equation (\ref{eq:phi_m}) with $\OmegaP=\Omega_\mathrm{vrt}$. The openness of the vortex-driven spirals would then naturally translate into a longer radial wave-length of the corresponding $\Texs$ oscillations. 

However, as mentioned earlier in Section \ref{sec:torque-analysis}, one has to remember that the long-term (time-averaged) behaviour of $\Texs$ (which is what is shown in Fig. \ref{fig:binary_dTdr_N}) is determined only by the component of $\delta\Sigma$ that is stationary in the frame co-rotating with the binary. At the same time, since $\Omega_\mathrm{vrt}\neq\Omegab$, the vortices, as well as the spiral pattern that they excite, are not stationary in that frame. As a result, the net contribution of the vortex-driven modes to $\delta\Sigma$ in the binary frame should average out to zero over long time intervals. We illustrate this expectation in Fig. \ref{fig:time-average}, where we plot both the snapshot of $\delta\Sigma$ at $t=50\Pb$ as well as the time average (over 10 $\Pb$ and centred on that $t$) of $\delta\Sigma$ in the frame co-rotating with the binary (it is also instructive to compare with Figs.  \ref{fig:polar_sig}h \& \ref{fig:dsig_dvort_map}e,j). This calculation is done for $\qb=1$, for which the vortices are strongest (see Fig. \ref{fig:dsig_dvort_map}) and clearly illustrates the dramatic effect of time-averaging: it very effectively filters out the vortex-driven perturbation at radii where the local orbital period is shorter than the averaging timescale (one can still see some traces of the vortex-driven modes in panel (b) outside $\approx 5\ab$, where this condition is not fulfilled). As a result, the time-averaged $\delta\Sigma$ retains only the structure due to the binary-driven $m=2$ modes, which are tightly wrapped. We also verified that for $\qb=10^{-3}$ time averaging does not change $\delta\Sigma$ pattern at $t=50\Pb$ as vortices play minimal role for such low $\qb$ at that time.

In the end of this exercise,  when we compute $\Texs$ and take its mean over time longer than the local orbital period, the contribution of the vortex-driven modes to the gravitational torque averages out to zero. Thus, the left panels of the Fig. \ref{fig:binary_dTdr_N} show us $\langle\Texs\rangle$ predominantly due to the tightly-wrapped $m=2$ binary-driven spirals, with no contribution from the vortex-driven modes in the inner disc, where the averaging interval is longer than the local orbital period. This naturally explains good agreement that we observe in Figure \ref{fig:spacing_zeros_dtdr_var_q} between the simulation data for very different $\qb$ and the WKB prediction (\ref{eq:dR_k}).

\begin{figure*}
	\begin{center}
	\includegraphics[width=\textwidth]{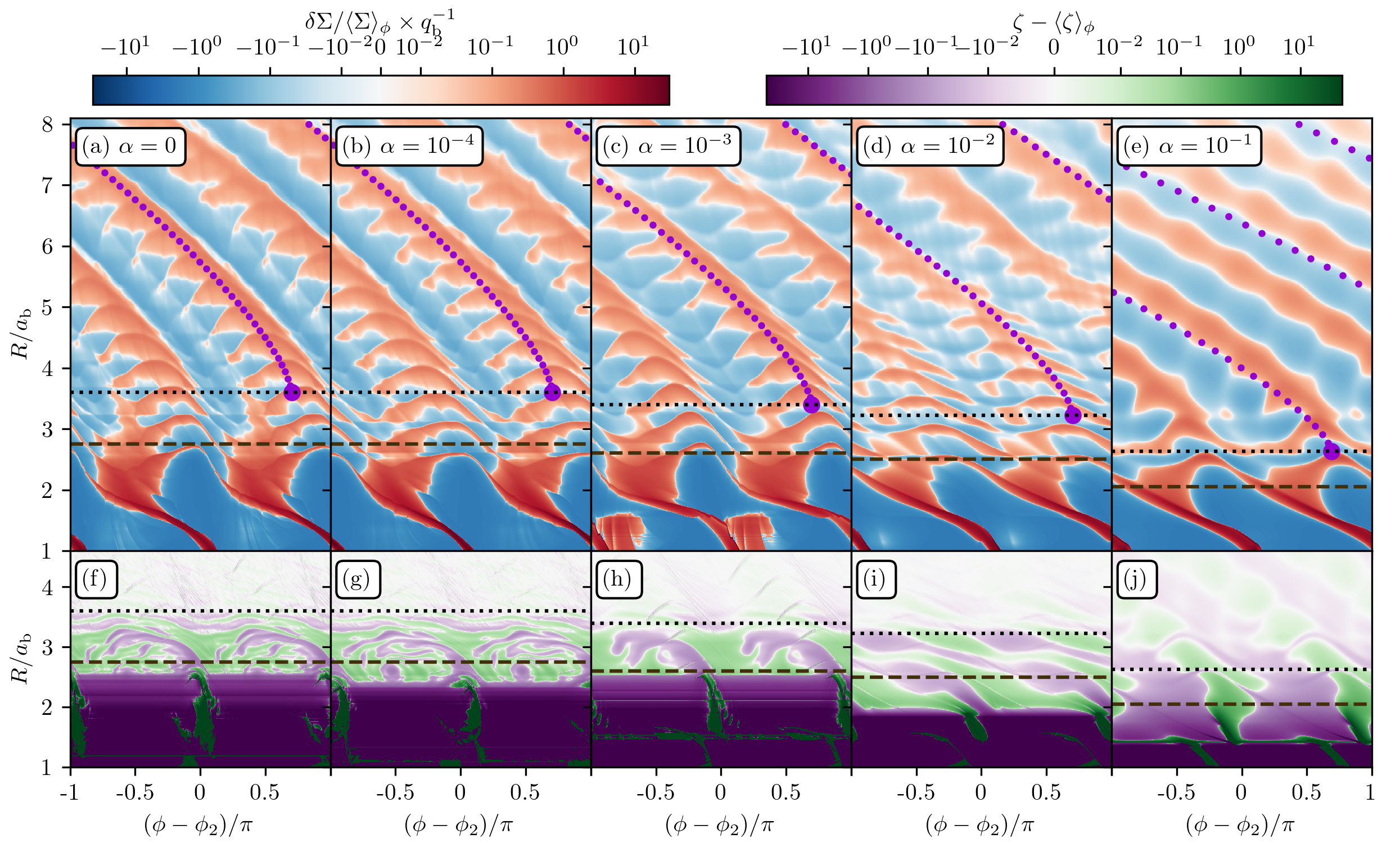}
	\caption{
		Same as Fig. \ref{fig:dsig_dvort_map}, for $\qb = 1$ at $t = 50\Pb$ and varying viscosity ($\alpha$ increasing left to right, indicated in panels). We use $m=2$ for all WKB fits  (\ref{eq:phi_m}), (\ref{eq:pitch}) and $\OmegaP$ = 0.22 $\Omegab$, 0.22 $\Omegab$, 0.24 $\Omegab$, 0.26 $\Omegab$, 0.36 $\Omegab$ from left to right, respectively.
 }
	\label{fig:visc_dsig_dvort_map_t50}
	\end{center}
\end{figure*}

\begin{figure*}
	\begin{center}
	\includegraphics[width=\textwidth]{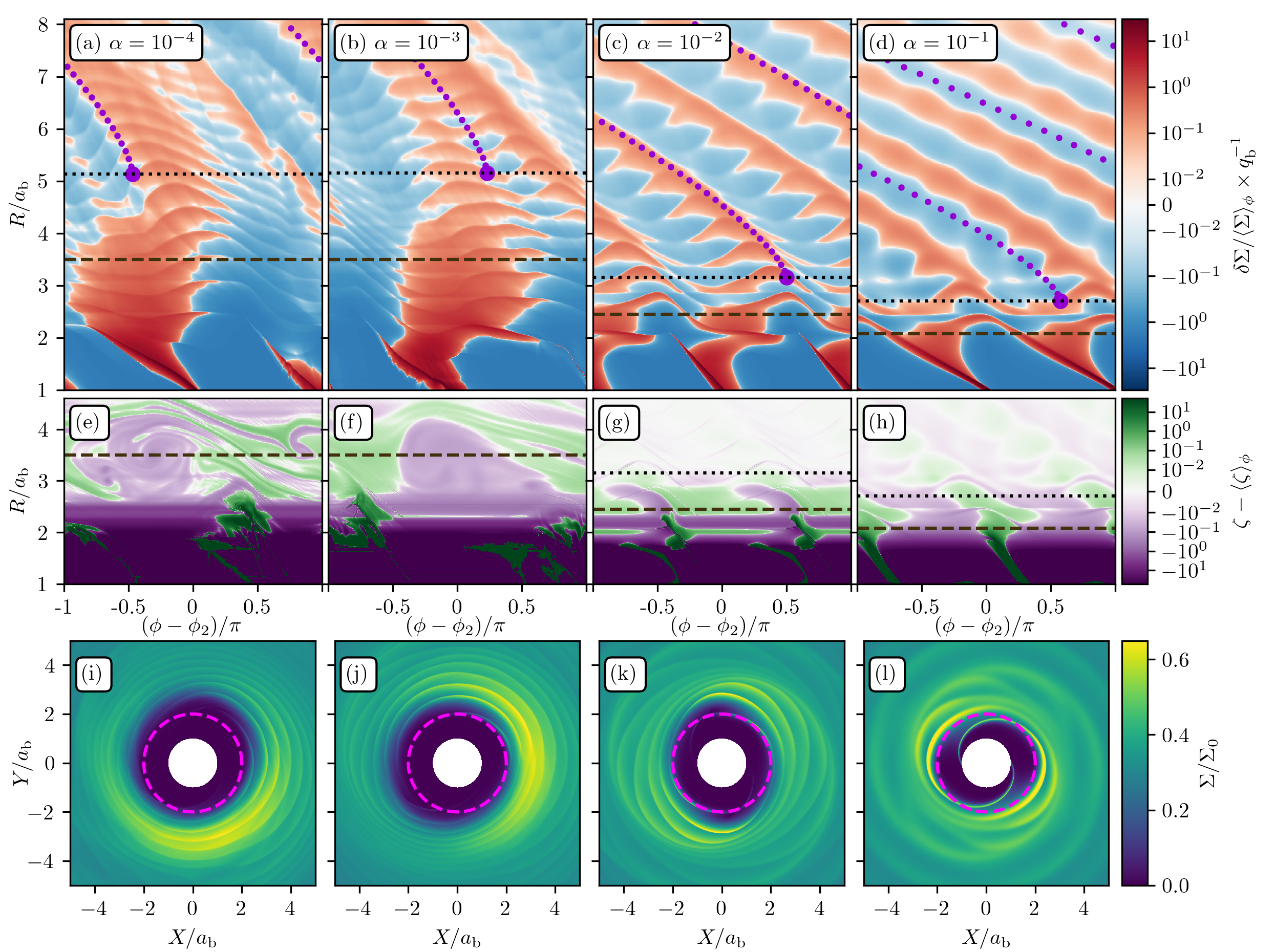}
	\caption{
		Same as Fig. \ref{fig:visc_dsig_dvort_map_t50} but at $t = 300 \Pb$ and only showing viscous models. 
  We use $m=1$ for the first two (a,b) and $m=2$ for the second two (c,d) WKB fits (\ref{eq:phi_m}), (\ref{eq:pitch}), with pattern speeds
  $\OmegaP = 0.17\Omegab,0.17\Omegab,0.26\Omegab,0.34\Omegab$, from left to right, respectively. The bottom row of this figure illustrates the shape of the central cavity in polar coordinates in each of these runs at late times.  The pink dashed circles have a radius of $R_\mathrm{c} = 2\ab$, and provide a reference for the shape of the cavity.
 }
	\label{fig:visc_dsig_dvort_map_t300}
	\end{center}
\end{figure*}


\subsection{Origin of the vortices}
\label{sec:orig_vort}


While vortices are not the main subject of this study, we still provide some discussion of their possible origin. 

In our simulations CBD starts with a sharp vortensity gradient in its inner region simply because of the sharp variation of $\Sigma(R,t=0)$ at the edge of the cavity, see Eq. (\ref{eq:Sigma0}). It is known \citep{Ono2016,Ono2018} that a radially localized and large-amplitude vortensity feature can be a natural driver of the Rossby Wave Instability \citep[RWI, see][]{Lovelace1999}. As RWI eventually results in non-axisymmetric vorticity concentrations, it represents a natural candidate for producing near-cavity vortices in our globally isothermal setting. 

However, RWI driven by the initial vortensity structure in the disc is not the whole story: if it were the case, vortices would be produced in the same way for all values of $\qb$ in Fig. \ref{fig:dsig_dvort_map} since the initial $\Sigma(R,t=0)$ is the same. Instead, there is a clear dependence on $\qb$, with vortices appearing earlier in higher-$\qb$ systems in Fig. \ref{fig:dsig_dvort_map}. Thus, binary mass ratio is an important parameter determining the conditions for the RWI (or some other instability) to be triggered. 

The most likely reason for this $\qb$-dependence is the nonlinear dissipation of the density waves launched by the binary in the disc: the higher $\qb$ is, the more nonlinear are the density waves (see Fig. \ref{fig:fourier_dsig}), leading to faster production of vortensity at the wave shocks near the cavity and ultimately causing the RWI to be triggered earlier. This picture is similar to the excitation of vortices by the RWI at the edges of planet-induced gaps in protoplanetary discs \citep{Koller2003,dvB2007,Lin2010}, with the more massive planets driving stronger density waves, producing vortensity via shock dissipation faster and causing earlier triggering of the RWI and vortex formation \citep{Cimerman2023a}.

At the same time, the initial vortensity gradient present near the cavity edge likely also plays an important role, shortening the time for vortices to appear. In that regard, the CBD situation bears similarity also with the production of vortices near the boundary layers of accretion discs as described in \citet{Coleman2022a}, in which a localised vortensity structure is present from the start because of the sharp gradient of the angular velocity in the boundary layer, see Fig. 8 of that work. In the planetary gap context such a vortensity feature is initially absent and arises only due to the action of the planet \citep{Cimerman2023a,RC2023}.

To summarize, the vortices in CBDs are likely produced by the RWI triggered by the evolving vortensity structure near the cavity, which gets established during the initial cavity clearing phase and then via a steady vortensity buildup due to the dissipation of the binary-driven density waves. Importantly, regardless of the origin of the vortices, we showed that the vortex-driven modes observed in our present work are identical to the vortex-driven modes excited in the boundary layers of accretion discs \citep{Coleman2022a} and at the edges of the planet-induced gaps \citep{Cimerman2023a}.


\subsection{Vortices and lumps in CBDs}
\label{sec:lumps}


On timescales much longer than $\Pb$ the CBD is often observed to develop a non-axisymmetric, eccentric cavity \citep{MacFadyen2008}. Such a cavity slowly precesses in the binary frame and typically features an overdensity --- a lump --- near its apocentre \citep{Shi2012}. This lump has an effect on the periodicity of accretion and is often associated with the strong $m=1$ spiral in the CBD.

Recently, using 2D inviscid general-relativistic hydrodynamical simulations \citet{Mignon2023} have followed the emergence of a lump at the inner edge of the disc. They ascribed its origin to the RWI that naturally develops at the inner edge of the disc and produces large-scale vortices that eventually evolve into a lump. Formation of a large vortex was suggested in that work as the root cause of the perturbation symmetry transitioning from $m=2$ to $m=1$ in circular equal mass binaries, followed by the formation of an eccentric cavity. \citet{Rabago2023} also found vortices forming at the inner edge of a polar disc around an eccentric binary in their low-viscosity runs.

Our results on the emergence of vortices and their role in the angular momentum transport support these findings and demonstrate that neither general relativity nor high disc inclination are necessary for the vortices to appear. At the same time, our findings do not provide a clear evidence in favor or against vortices being a precursor to and the cause for the lump formation. As an alternative, \citet{Shi2012} and \citet{Noble2012} have ascribed the lump origin to the interaction of gas streams inside the cavity with the cavity edge.   

One way to discriminate between the different possibilities might be to examine the sensitivity of vorticity structures in CBDs to the disc viscosity, since viscous stresses are known to suppress the RWI while lumps have been observed even in high-viscosity simulations. We provide a quick look into this issue in Fig. \ref{fig:visc_dsig_dvort_map_t50}, where we show snapshots of $\delta\Sigma$ and the non-axisymmetric component of vortensity for different values of viscosity (parameterized via an effective $\alpha \in \lbrace 10^{-4}, 10^{-3}, 10^{-2}, 10^{-1} \rbrace$, see \citealt{Shakura1973}) in a circular equal-mass binary case. The snapshots are taken at $t=50\Pb$, same as in Fig. \ref{fig:dsig_dvort_map} (right panels of Fig. \ref{fig:dsig_dvort_map} are identical to the left panels in Fig. \ref{fig:visc_dsig_dvort_map_t50}). One can see that vortices readily form for $\alpha\le 10^{-3}$, and the disc and vortensity structure are almost the same in these runs --- two strong rolls around $2.7\ab$ launching open spiral arms into the disc, see three left panels. In higher viscosity ($\alpha\ge 10^{-2}$) runs $\zeta$ structure changes and while there is still a clear $m=2$ periodicity in the vortensity maps, the characteristic roll structures are no longer there. The two spiral arms are still present in $\delta\Sigma$ maps, albeit less open than for lower $\alpha$ (most likely because high viscosity reduces the size of the cavity and moves the vortensity structures closer to the binary, increasing their $\OmegaP$ and decreasing their pitch angle, see Eq. (\ref{eq:pitch})). 

In Fig. \ref{fig:visc_dsig_dvort_map_t300} we illustrate the appearance of viscous CBDs at later time, $t=300\Pb$ (we only show $\alpha\ge 10^{-4}$ runs). One can see that in low-viscosity discs with $\alpha\le 10^{-3}$ a single dominant vortex forms by that time at $R\sim (3-4)\ab$ launching a very open and wide spiral arm in the disc. But in the higher $\alpha\ge 10^{-2}$ runs the picture hardly changes compared to $t=50\Pb$ shown in Fig. \ref{fig:visc_dsig_dvort_map_t50}, with $m=2$ azimuthal periodicity still clearly dominating. This difference is also clearly reflected in the shape of the central cavity as illustrated in the bottom row of Fig. \ref{fig:visc_dsig_dvort_map_t300}: for low $\alpha=10^{-4},10^{-3}$ the cavity is eccentric with a well-defined lump at its edge, similar to the 'asymmetric' state of \citet{DOrazio2013}, whereas for higher $\alpha=10^{-2}, 0.1$ the cavity shape conforms with the $m=2$ perturbation (ellipse centred on the binary barycentre), similar to the 'point-symmetric' state of \citet{DOrazio2013}. Thus, $m=2$ symmetry breaking is clearly present in our low-$\alpha$ discs, which exhibit vortices early on and feature a strong vortex at $t=300\Pb$, but not in high-$\alpha$ CBDs, at least during the run time of our viscous simulations (which is still quite short compared to many other CBD simulations evolved for thousands of orbits, e.g. \citealt{Miranda2017,Siwek2023}).

In conclusion of this section we also note that in all our viscous runs one can still easily see the interference of the open vortex-driven waves with the tightly-wound binary-driven waves, manifested in the form of ripples on top of the vortex-driven spiral arms. Very importantly, the oscillations of $\Texs$ in viscous discs remain consistent with the explanations provided in Sections \ref{sec:torque-origin}, \ref{sec:torque-analysis}.


\section{Discussion}
\label{sec:discuss}


Our results on $\Texs$ oscillations and their interpretation imply that the main agent determining the radial structure and amplitude of the oscillations is the behaviour of the density waves launched by the binary into the CBD with the pattern speed $\OmegaP=\Omegab$. These waves get excited close to the edge of the cavity and then freely propagate through the disc, losing their strength as they dissipate due to non-linear (or linear) processes. This implies that the torque oscillations owe their existence to {\it coupling of the binary potential to the freely-propagating density waves} in the disc. 

The most obvious driver of the density waves with $\OmegaP=\Omegab$ in the CBD is the resonant wave excitation by the binary potential at the low order Lindblad resonances \citep{GT79,GT80,Art1994}. But in the case of a CBD some contribution to the wave excitation may also come from the ballistic motion of the gas inside the cavity, with fluid splashback against the cavity wall exciting nontrivial response in the CBD \citep{Farris2014}. However, in our picture the actual mechanism driving the density waves in the CBD
does not really matter   --- once the waves exist the torque density will exhibit oscillatory structure. 

Note that, as mentioned in Section \ref{sec:vortices}, vortices forming near the cavity edge do not contribute to the gravitational $\Texs$ oscillations in the time-averaged sense since their $\OmegaP\neq\Omegab$ and their density waves are launched non-gravitationally. While vortices do complicate our interpretation of $\Texs$ behaviour, a careful analysis allows us to isolate their effect on the angular momentum transport in the CBD, see Section \ref{sec:vortices}.
 
\begin{figure}
	\begin{center}
	\includegraphics[width=0.49\textwidth]{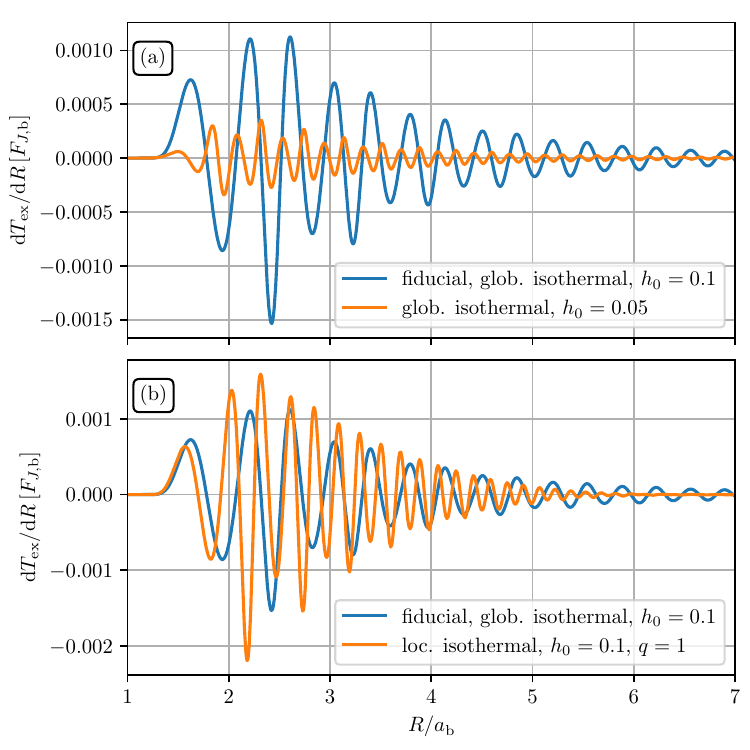}
	\caption{
    Comparison of the $\Texs$ profiles from our fiducial globally isothermal ($q=0$) runs with $h_0=0.1$ (blue) and simulations with other thermodynamic assumptions (orange); all runs are for $\qb=10^{-3}$, as in Fig. \ref{fig:Airy}. (a) Same EoS but $h_0=0.05$, which naturally leads to doubling of the radial frequency of $\Texs$ oscillations, consistent with Eqs. (\ref{eq:dR_k}), (\ref{eq:dR_k1}). (b) Same $h_0=0.1$ but a different EoS --- locally isothermal with $q=1$, i.e. $T\propto R^{-1}$. As $\cs$ decreases with $R$ for the orange curve, the radial period of $\Texs$ oscillations also decreases, consistent with the equation (\ref{eq:dR_k1}).
	}
	\label{fig:comp_dtdr_var_h_var_eos_comb}
	\end{center}
\end{figure}

\begin{figure}
	\begin{center}
	\includegraphics[width=0.49\textwidth]{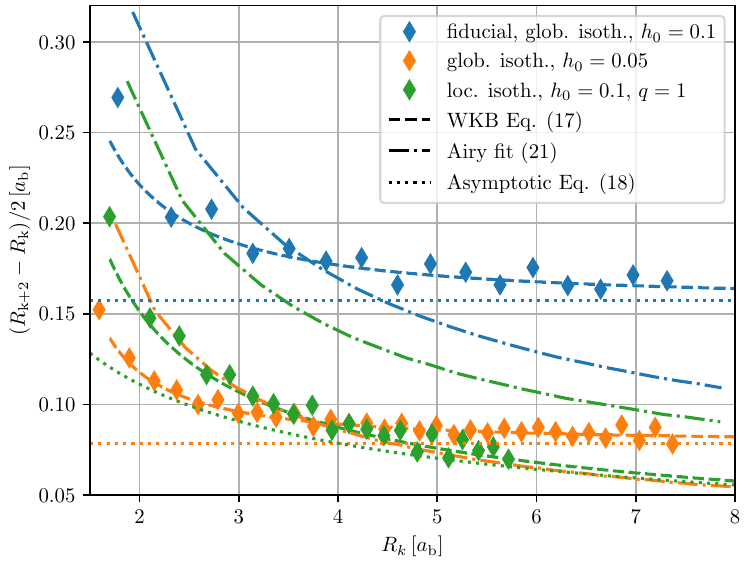}
	\caption{
    Illustration of the radial periodicity of $\Texs$ oscillations (similar to Fig. \ref{fig:spacing_zeros_dtdr_var_q}) for different thermodynamic assumptions: globally isothermal EoS with $h_0=0.1$ (blue) and $h_0=0.05$ (orange) and locally isothermal EoS with $q=1$ and $h_0=0.1$ (green). Coloured points show the corresponding simulation data for the distance between $\Texs$ nodes, dashed and dotted curves illustrate WKB equations (\ref{eq:dR_k}) and (\ref{eq:dR_k1}), respectively, while the dot-dashed curves show the resonant torque prediction (\ref{eq:dR_k-res}).
 }
	\label{fig:zeros_spacings}
	\end{center}
\end{figure}

The primary goal of this study was to understand the spatial structure of the $\Texs$ oscillations, namely their {\it radial quasi-periodicity}. Equations (\ref{eq:dR_k}), (\ref{eq:dR_k1}) show that this periodicity depends almost exclusively on the radial behaviour of the sound speed $\cs$, at least in the linear regime. 
While all the simulations that we presented so far assume globally isothermal EoS and $h_0=0.1$, this remains true for other thermodynamic assumptions as well. To show this, in Fig. \ref{fig:comp_dtdr_var_h_var_eos_comb} we show $\Texs$ oscillations for $\qb=10^{-3}$ but different thermodynamic setups: a lower $h_0=0.05$, panel (a), and a different (locally isothermal with $q=1$) EoS in panel (b). One can see that in both cases the radial periodicity of $\Texs$ behaves in agreement with Eqs.
(\ref{eq:dR_k}), (\ref{eq:dR_k1}): the (constant) radial period goes down by a factor of 2 compared to the fiducial case in panel (a), while in panel (b) the radial period decreases with $R$. This radial period evolution is in full agreement with our theory as we additionally illustrate in Fig. \ref{fig:zeros_spacings}. There we plot (similar to Fig. \ref{fig:spacing_zeros_dtdr_var_q}) the radial spacing between the $\Texs$ nodes (as a function of $R$) from simulations with different thermodynamic assumptions and $\qb=10^{-3}$. One can see that in each case the simulation data points (diamonds) agree very well with the analytical WKB prediction 
(\ref{eq:dR_k}), asymptotically converging to the behaviour given by Eq. (\ref{eq:dR_k1}). Note also that the insensitivity of our theoretical calculations in Section \ref{sec:torque-analysis} to the behaviour of the background surface density $\azav{\Sigma}$ makes our simulations and theoretical results immune to any possible inconsistencies in setting up the initial disc profile in Section \ref{sec:ICs}, e.g. for low $\qb$.

We have also linked the evolution of the amplitude of the $\Texs$ oscillations to that of the individual Fourier harmonics of the CBD perturbation $\delta\Sigma_m$, see Section \ref{sec:torque-analysis} and Eq. (\ref{eq:Texm}). However, understanding the radial behaviour of $\delta\Sigma_m$ is beyond the scope of this study. Some information on the evolution of the wave amplitude due to the nonlinear effects can be gleaned from  Fig. \ref{fig:fourier_dsig}, but the full quantitative understanding of these results, including the prediction of the overall density wave amplitude as a function of $\qb$ and other parameters, will require deeper understanding of the wave excitation and $\azav{\Sigma}$ structure near the cavity (going beyond the simple fit (\ref{eq:Sigma0})) in the first place. Wave amplitude evolution will also be affected by the thermodynamic assumptions about the equation of state \citep{Miranda2019II} and explicit cooling mechanisms \citep{Miranda2020I,Miranda2020II} in the CBD.

In this work torque density oscillations have been studied in the context of a CBD with a large cavity in which the masses perturbing the disc orbit. A similar phenomenon of the torque wiggles has been explored by \citet{Cimerman2023b} in smooth (i.e. not featuring gaps) protoplanetary discs in the context of disc-planet interaction, see Section \ref{sec:planet}. But other astrophysical discs may also feature $\Texs$ oscillations of similar nature. 

For example, in their study of discs in cataclysmic variables (CVs) orbiting one component of the binary \citet{Ju2016} reported oscillations of $\Texs$ (see their Fig. 10) very similar to what we and others find in CBDs; even earlier such oscillations have been reported by \citet{Sav1994}. Based on our current understanding, we believe that the torque density oscillations in CV discs are caused by the coupling of the gravitational potential of the secondary to the freely (inward) propagating spiral density waves launched by the secondary at the outer edge of the disc; these waves are clearly visible in Figs. 5 and 7 of \citet{Ju2016}. We leave detailed exploration of torque oscillations in CV discs to a future study.


\subsection{Resonant torques and $\Texs$ oscillations}
\label{sec:resonant-torque}


Until this work, the standard explanation for the $\Texs$ oscillations relied on the $m=2$ eigenfunction of the resonant torque density excited at the corresponding outer Lindblad resonance where $\Omega_2=(2/3)\Omegab$ \citep{MacFadyen2008,Farris2011,Shi2012}. The general idea ascends to \citet{Meyer1987} who showed that an isolated $m=2$ Lindblad resonance gives rise to a torque density in the form
\begin{align}
\frac{\de T_2}{\de R}=
K_2\,\mathrm{Ai}\left(\frac{R_2-R}{\lambda_2}\right),
\label{eq:res-T}
\end{align}
where  $R_2$ is the OLR radius for $m=2$, $\lambda_2=2^{-2/3} (H/R)_2^{2/3}\ab$ and $(H/R)_2$ is the disc aspect ratio evaluated in the vicinity of $R_2$. The dimensional, radially-independent coefficient\footnote{The explicit expression for its approximate value is given by Eq. (31) in \citet{MacFadyen2008}.} $K_2$ is proportional to $\Sigma(R_2)$, which makes the determination of its exact value rather ambiguous because of the rapid variation of $\Sigma$ near the resonance. For that reason, in Fig. \ref{fig:Airy} we simply choose $K_2$ so that $\de T_2/\de R$ approximately matches the first peak of $\Texs$.

Given the asymptotic behaviour of the Airy function $\mathrm{Ai}(-z)\propto z^{-1/4}\sin\left[(2/3)z^{3/2}+\xi_0\right]$, $\xi_0=$ const, for $z\gg 1$ \citep{Abra1972}, Eq. (\ref{eq:res-T}) predicts the distance between the consecutive nodes of $\de T_2/\de R$ to obey
\begin{align}
 R_{k+1}-R_k \approx \frac{\ab}{2}(H/R)_2\,\sqrt{\frac{\ab}{R_k}},~~~~~~R_k\gg \ab.
 \label{eq:dR_k-res}
\end{align}
This expression is clearly different from our prediction (\ref{eq:dR_k1}) since the dependence of $\cs(R)$ in the latter can be arbitrary. In particular, globally isothermal simulations presented in this work have $\cs(R)=$ const and the scaling of $R_{k+1}-R_k$ with $R_k\gg \ab$ in (\ref{eq:dR_k1}) is different from that in (\ref{eq:dR_k-res}), as illustrated in Fig. \ref{fig:Airy}. This is also very obvious in Figs.   \ref{fig:spacing_zeros_dtdr_var_q} and \ref{fig:zeros_spacings}, where the dot-dashed curves representing Eq. (\ref{eq:dR_k-res}) provide an inadequate fit to the simulation points for different thermodynamic assumptions, unlike our WKB predictions (\ref{eq:dR_k}), (\ref{eq:dR_k1}). 

Moreover, scaling of the radial period of $\Texs$ with $R$ is only a part of the story. For example, in a locally-isothermal disc with $\cs(R)\propto R^{-1/2}$ (i.e. $q=1$ and a radially-independent aspect ratio $H/R$) Eq. (\ref{eq:dR_k}) predicts $R_{k+1}-R_k\propto R_k^{-1/2}$, just as in Eq. (\ref{eq:dR_k-res}). However, the two predictions are still off by a factor $\approx 2.2$, see green dot-dashed and dotted curves for $q=1$ in Fig. \ref{fig:zeros_spacings}. Also, Eq. (\ref{eq:res-T}) implies that for $R\gg \ab$ the amplitude of $\de T_2/\de R \propto R^{-1/4}$, whereas our result (\ref{eq:Texm}) for the behaviour of torque oscillations means that their amplitude should scale as $R^{-2}|\delta\Sigma_2(R)|$ (see Section \ref{sec:torque-analysis}), i.e. a very different dependence. 

These observations and a poor match that $\de T_2/\de R$ provides for $\Texs$ in Figs.  \ref{fig:Airy}, \ref{fig:spacing_zeros_dtdr_var_q} \& \ref{fig:zeros_spacings} make it clear that the resonant torques cannot explain the behaviour of the torque oscillations. The reason for this naturally follows from the original work of \citet{Meyer1987}, which analyzed density wave excitation only in the {\it immediate vicinity of a Lindblad resonance}. The Airy function dependence arises in the behaviour of the velocity perturbation when fluid equations are expanded near $R_2$ and it propagates also into $\de T_2/\de R$. However, by construction, Eq. (\ref{eq:res-T}) is then also valid only locally, very close to $R_2$ and is simply not applicable for $R-R_2\gtrsim \ab$ where the conspicuous torque oscillations are found. This invalidates the idea that the mathematical structure of the resonant torque expression (\ref{eq:res-T}) can be the reason for the global torque oscillations. The only resonance relevant for the origin of these oscillations is the corotation of the binary and the wave pattern that it drives in the disc, since only the component of $\delta\Sigma$ stationary in the binary frame is responsible for the mean $\Texs$, see Sections \ref{sec:torque-analysis}, \ref{sec:vortices}.

However, resonant torques still play a role in the genesis of $\Texs$ oscillations, even though indirectly: they are the primary cause of the density wave excitation in the first place. It is the subsequent coupling of the binary potential to the freely-propagating density waves that causes the oscillatory $\Texs$ as we described in Sections \ref{sec:torque-origin} \& \ref{sec:torque-analysis}. The resonant torque endows density waves with the angular momentum that they carry away into the disc, and most of this angular momentum gets accumulated very close to the resonance \citep{Meyer1987}. Far from the resonance, the excitation torque density $\Texs$ provides an insignificant contribution to the integrated torque $T_\mathrm{ex}$ due to its oscillatory behaviour, as can be seen in the right panels of Fig. \ref{fig:binary_dTdr_N}.


\subsection{Comparison with the planetary case}
\label{sec:planet}


\begin{figure}
	\begin{center}
	\includegraphics[width=0.49\textwidth]{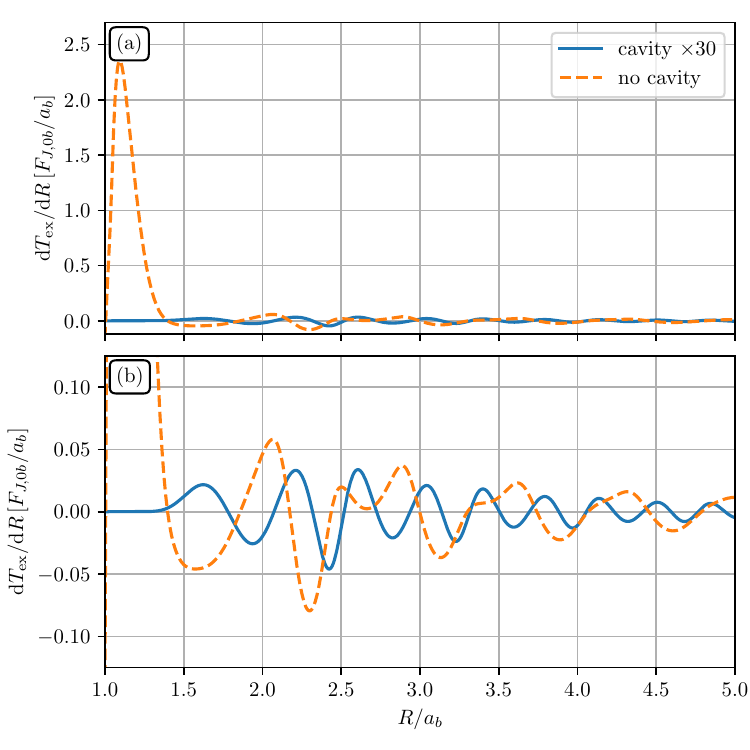}
	\caption{
    Comparison of the excitation torque density profiles (on different scales, top vs botom) in the circumbinary setup, with central cavity (blue) and planetary setup, with no cavity or gap around the planetary orbit (orange). Both are obtained from the corresponding simulations with $\qb=10^{-3}$. In the CBD case $\Texs$ has been multiplied by $30$ to bring the scale of its oscillations to the level of the torque wiggles in the planetary case (see bottom panel). The top panel illustrates the tall, main peak of $\Texs$ near the planet, which is absent in the CBD case. Note the quasi-periodicity of the torque wiggles (orange), which has a radial length scale twice larger than that of the $\Texs$ oscillations in the CBD (blue).
	}
	\label{fig:comp_dtdr_cav_nocav}
	\end{center}
\end{figure}

Recently \citet{Cimerman2023b} described and explained the phenomenon of the so-called torque wiggles in disc-planet interaction. They showed that when a planet is embedded in a smooth disc (without a gap), the torque density profile far from the planet shows a quasi-periodic pattern of low-amplitude $\Texs$ oscillations which are especially coherent in the outer disc. The explanation that was provided for these torque wiggles in \citet{Cimerman2023b}  --- coupling of the planetary potential to the freely-propagating density wave launched by the planet --- is essentially the same as the explanation of $\Texs$ oscillations in our present work and in fact motivates it. 

It is instructive to provide a side-by-side comparison of the torque oscillations and the torque wiggles, which we do in Fig. \ref{fig:comp_dtdr_cav_nocav} by showing $\Texs$ profiles extracted from two simulations with $\qb=10^{-3}$. The first one is our standard CBD simulation with a surface density profile in the form (\ref{eq:Sigma0}) featuring a central cavity (blue curves). The other one assumes a planetary setup, in which the secondary is embedded in a smooth disc, i.e. there are no radial structures around the orbit of the secondary (orange curves). Given that this $\qb\sim (H/R)^3$ is at the boundary of the linear regime, one expects gap opening to eventually occur in the latter simulation. However, we ran it for only a short period of time sufficient to measure $\Texs$ while the disc was not strongly perturbed.  

The first obvious thing to note in Fig. \ref{fig:comp_dtdr_cav_nocav} is the very different scale of the $\Texs$ in two cases --- note that $\Texs$ in CBD has been multiplied by a factor of 30 to make it similar in scale to the torque wiggles in the outer disc. The second remarkable feature is the strong peak of $\Texs$ near the planetary orbit in the planetary setup, see panel (a), which is completely missing in the CBD case.  Both of these differences are due to the lack of gas near the perturber in the CBD setup, which does not allow the high-order Lindblad resonances to be activated. It is well known \citep{GT80} that in the planetary setting most angular momentum is carried by the wave harmonics with azimuthal wavenumbers $m\sim R/h\gg 1$, which are excited close to the planet and explain the strong peak of $\Texs$. Because of this peak, the relative amplitude of the torque wiggles in the planetary case is low, at the level of several per cent of the maximum $|\Texs|$ \citep{Cimerman2023b}. On the contrary, in the CBD case,  because of the central cavity,  only the $m=2$ outer Lindblad resonance couples to the disc sufficiently well to drive this low-order harmonic. Since the central peak is absent in this case, the torque oscillations have $O(1)$ amplitude and are much more prominent. 

Focusing next on the radial structure of $\Texs$ far from the planet in panel (b), one can see that outside the 'negative torque density' region \citep[at $R\in (1.4,1.8)\ab$, see][]{Dong2011,RP12}  planetary $\Texs$ shows a roughly self-similar periodic structure with two peaks and a deeper trough, with slowly decaying amplitude. At the same time, the CBD $\Texs$ has an almost sinusoidal shape with a modulated amplitude and the radial frequency which is twice that of the planetary $\Texs$ pattern. The explanation of these differences lies in the different spatial structure of the density waves in two cases: in the planetary setup there is a narrow, single-armed density wave \citep{R02,OL02} which is a superposition of many azimuthal harmonics but is dominated by high $m\sim R/h$. As shown by \citet{Cimerman2023b}, this results in a nontrivial shape of the torque wiggles. In the CBD case a low $m=2$ harmonic dominates (see Fig. \ref{fig:fourier_dsig}) since the cavity suppresses higher $m$, ensuring a single dominant sinusoidal contribution to $\Texs$ as follows from Eq. (\ref{eq:Texm}). The radial frequency is twice higher in the CBD case because the density wave has two arms rather than one as in the planetary case --- a similar (although not exactly the same) situation can be seen in Fig. \ref{fig:toy_comb} for $m=1$ and $m=2$.  

Despite these differences, the underlying cause of the quasi-periodicity of the torque wiggles and $\Texs$ oscillations is the same ---  the presence of the density wake in the disc and its coupling to the perturber's potential. Since the wake shape is set mainly by the radial profile of the sound speed $\cs$, in both cases $\azav{\Sigma(R)}$ profile plays essentially no role in setting the radial periodicity of $\Texs$.


\subsection{Comparison to previous studies}
\label{sec:comp_prev}


As mentioned in the Introduction, a number of CBD studies have previously looked into the behaviour of $\Texs$. These investigations used both 2D and 3D setups, considered standard $\alpha$ viscosity as well as full MHD models allowing for magnetorotational instability, and some of them included general relativistic effects. Despite these differences, all these studies confirm the large-amplitude, oscillatory nature of $\Texs$ in CBDs.  

Unlike most other studies using explicit shear viscosity, we mainly considered the inviscid limit. This makes our work similar to \citet{Mignon2023} and \citet{Rabago2023} who also explored low-viscosity CBDs and found, similar to us, that vortices readily form at the edge of the cavity, see Section \ref{sec:vortices}. Compared to many other studies \citep[e.g. ][]{Siwek2023}, our simulations are run for a rather short time. Nevertheless, our results will be helpful for understanding the outcomes of the longer-term simulations. 

Working in the planetary setting, \citet{Dempsey2020} presented results on the evolution of $\Texs$ as the width of the gap carved by the planet increases. By examining their Fig. 5 we see a gradual transition from the radially-spaced, low-amplitude torque wiggles in the case of a narrow and shallow gap, still allowing wave excitation at high-$m$ Lindblad resonances, to the less radially-spaced, large-amplitude torque density oscillations in the case of a deep and wide gap, similar to a cavity in the CBD. This evolution can be easily understood based on our discussion in Section \ref{sec:planet}: as the gap depth and width increase, higher-$m$ perturbation harmonics get gradually deactivated, until $m=2$ becomes the dominant one. As a result, the relative amplitude of  of $\Texs$ oscillations increases (as the $\Texs$ peak near the planet is gradually eroded) and the one-armed spiral perturbation transitions to a two-armed density wave, leading to the increased radial frequency of oscillations. 

In conclusion, we comment on a couple of byproducts of our work that may be useful for interpreting the results of other studies. In their investigation of CBD dynamics \citet{Mahesh2023} presented Fourier decomposition of $\delta\Sigma$ (see their Fig. 8) and argued, based on the rough radial independence of the relative contribution of different harmonics, that the resonant torques are not involved in setting the cavity size in the course of disc-binary interaction. Our calculation of $|\delta\Sigma_m|$ in Fig. \ref{fig:fourier_dsig} shows a very similar picture, with different harmonics maintaining roughly the same order as $R$ varies. But our interpretation is very different: this decomposition only reflects the harmonic content of the freely-propagating density waves launched by the binary (and not the radial scaling of different multipoles of the binary potential at $R\gg \ab$). And resonant torques are the reason why these waves are excited in the first place, they just have little to do with the radial structure of $\Texs$ oscillations.  Thus, we would caution against dismissing them in the CBD context.

In their study of waves in CV discs \citet{Van2023} found the emergence of a one-armed spiral arm in the inner disc and the trend towards the dominance of $m=1$ perturbation component in high Mach number (low $\cs$) discs. Based on our results on spiral arm merging in Figs.  \ref{fig:polar_sig}e and \ref{fig:dsig_dvort_map}b (see Sections \ref{sec:res-pert}, \ref{sec:torque-origin}), we suggest that in the CV context a single arm can also arise as a result of a nonlinearly driven merger of the two spiral arms excited at the outer edge of the CV disc. This tendency would get stronger as the flow Mach number and, correspondingly, the density wave nonlinearity increase, just as observed in \citet{Van2023}.


\section{Summary}
\label{sec:conc}


We have explored angular momentum exchange in a circumbinary disc and the properties of the excitation torque density $\Texs$ --- an important characteristic of the binary-disc gravitational coupling --- focusing on its nature and radial periodicity. Using 2D, primarily inviscid, hydrodynamic simulations and analytical arguments, we found the following.    

\begin{itemize}
    \item Outside the central cavity, in the bulk of the CBD (at $R\gtrsim \ab$), gravitational torque arises due to the coupling of the binary potential with the freely-propagating density waves launched near the cavity edge (similar to the origin of the torque wiggles explored by \citet{Cimerman2023b} in the planetary context).

    \item The radial periodicity of excitation torque oscillations is set primarily by the rate at which these density waves wrap around the origin, which in turn is determined by the gas sound speed.

    \item The periodicity of $\Texs$ also depends on the symmetry properties of the waves, i.e. a number of their spiral arms. Two-armed $m=2$ density waves often encountered in CBDs result in twice higher radial frequency of $\Texs$ oscillations than the $\Texs$ frequency for single-armed spirals.

    \item The evolution of the amplitude of $\Texs$ oscillations with radius is set by the density wave dissipation and the radial dependence of the binary potential.

    \item The local resonant Lindblad torque, often invoked to explain $\Texs$ oscillations, plays no direct role in their origin and periodicity. It is involved only indirectly, as the initial driver of the density waves that give rise to $\Texs$ oscillations.

    \item Angular momentum transport in the inner CBD can be affected in a non-trivial manner by the vortices that readily form (especially for high $\qb$) at the inner edge of the cavity in our inviscid simulations (although we observe their effect in our viscous runs as well). These vortices may be relevant for the origin of 'lumps' in CBDs.
    
\end{itemize}


To clarify the picture of the binary-CBD coupling, future work should aim to understand the overall amplitude and radial decay of the torque density in CBDs as a function of the binary mass ratio and disc parameters. Results of our study can also be used to understand the torque structure in other objects, e.g. in discs orbiting one of the components of the binary in cataclysmic variables, X-ray and young stellar binaries, see Section \ref{sec:discuss}.

\section*{Acknowledgements}


We are grateful to Ryan Miranda for useful discussions. N.P.C. is funded by an Isaac Newton Studentship and a Science and Technology Facilities Council (STFC) studentship.
R.R.R. acknowledges financial support through the Ambrose Monell Foundation, and STFC grant ST/T00049X/1. Part of this work was performed using resources provided by the Cambridge Service for Data Driven Discovery (CSD3) operated by the University of Cambridge Research Computing Service (\texttt{www.csd3.cam.ac.uk}), provided by Dell EMC and Intel using Tier-2 funding from the Engineering and Physical Sciences Research Council (capital grant EP/P020259/1), and DiRAC funding from the Science and Technology Facilities Council (\texttt{www.dirac.ac.uk}).

\noindent \textit{Software:} NumPy \citep{2020NumPy-Array}, SciPy \citep{2020SciPy-NMeth}, IPython \citep{IPython}, Matplotlib \citep{Matplotlib}, Athena++ \citep{Athenapp2020}.


\section*{Data Availability}
The data underlying this article will be shared on reasonable request to the corresponding author.




\bibliographystyle{mnras}
\bibliography{references} 




\appendix


\section{Analytical derivation for $\Texs$}
\label{app:derive}


We start by manipulating (\ref{eq:Texs}) and re-writing it more explicitly as
\begin{align}
\Tex = \frac{G}{R}\int\limits_0^{2\pi}\Sigma(R,\phi)\sum\limits_{i=1,2}\frac{M_i R_i\sin(\phi-\phi_i)\,\de \phi}{\left[1+\alpha_i^2-2\alpha_i\cos(\phi-\phi_i)\right]^{3/2}},
\label{eq:Texs1}
\end{align}
where $\alpha_i=R_i/R<1$. We then use the known identity \citep[e.g.][]{MurrayDermott}
\begin{align}
\frac{1}{\left(1+\alpha^2-2\alpha\cos\psi\right)^{1/2}} = \frac{1}{2}b_{1/2}^{(0)}(\alpha)
+\sum\limits_{n=1}^\infty b_{1/2}^{(n)}(\alpha)\cos (n\psi),
\label{eq:ident}
\end{align}
where
\begin{align}
    b^{(n)}_{1/2} (\alpha) = \frac{1}{\pi}
                    \int\limits_{0}^{2\pi}
                        \frac{\cos (n \psi)\,\de \psi}{(1 + \alpha^2 - 2 \alpha \cos \psi)^{1/2}},
    \label{eq:lap_def}
\end{align}
are the Laplace coefficients. Differentiating (\ref{eq:ident}) with respect to $\psi$, we find
\begin{align}
\frac{\sin\psi}{\left(1+\alpha^2-2\alpha\cos\psi\right)^{3/2}} = 
\alpha^{-1}\sum\limits_{n=1}^\infty nb_{1/2}^{(n)}(\alpha)\sin (n\psi).
\end{align}

This relation allows us to re-write (\ref{eq:Texs1}) as
\begin{align}
\Tex = G\int\limits_0^{2\pi}\Sigma(R,\phi)\sum\limits_{i=1,2}\sum\limits_{n=1}^\infty n M_i b^{(n)}_{1/2} (\alpha_i)\sin(n(\phi-\phi_i))\,\de \phi.
\label{eq:Texs2}
\end{align}
Swapping the orders of summations and integration and using the fact that $\phi_2=\phi_1+\pi$, we finally arrive at the relations (\ref{eq:Texs3})-(\ref{eq:Cn}).


\bsp	
\label{lastpage}
\end{document}